# Exact Distributions for the Solutions of the Compressible Viscous Navier Stokes Differential Equations: An Application in the Aeronautical Industry

Author's Name

Rensley Meulens

Author Affiliations
*Elux Technologies B.V.*
*Heintje Kool kv 183, Curaçao, Dutch Caribbean, 1000NA*
*Author Emails*
. Corresponding author: rensleymeulens@gmail.com)
The corresponding author is both the main author and the sole contributor to this manuscript.

**Abstract.** Wind tunnels and linearized turbulence and boundary-layer models have been so far necessary to simulate and approximate the stationery lift and drag forces on (base-mounted) airfoils by means of statistically determined or approximated values of the relevant situational coefficients as the drag and lift coefficients.To improve this process, we introduce transient and exact formulae to separate these forces in advance by means of the solutions found from the fluid dynamics model of the Navier Stokes differential equations.

## I INTRODUCTION: AIRFOIL ENGINEERING

The flow over an airfoil is determined by the pressure stresses and shear stresses working on the shear layer around the airfoil. The stresses are depended on the relative velocity with regard to the fluid dynamical medium wherein the airfoil resides, the angle of attack and the airfoil design. The flow is comparable with a flow over a driven-lid cavity as thorough discussed in the main article. In Fig. 1 are conventional airfoils [1]with their properties listed. The properties are calculated via wind-tunnel setups and or turbulence simulation models [2]

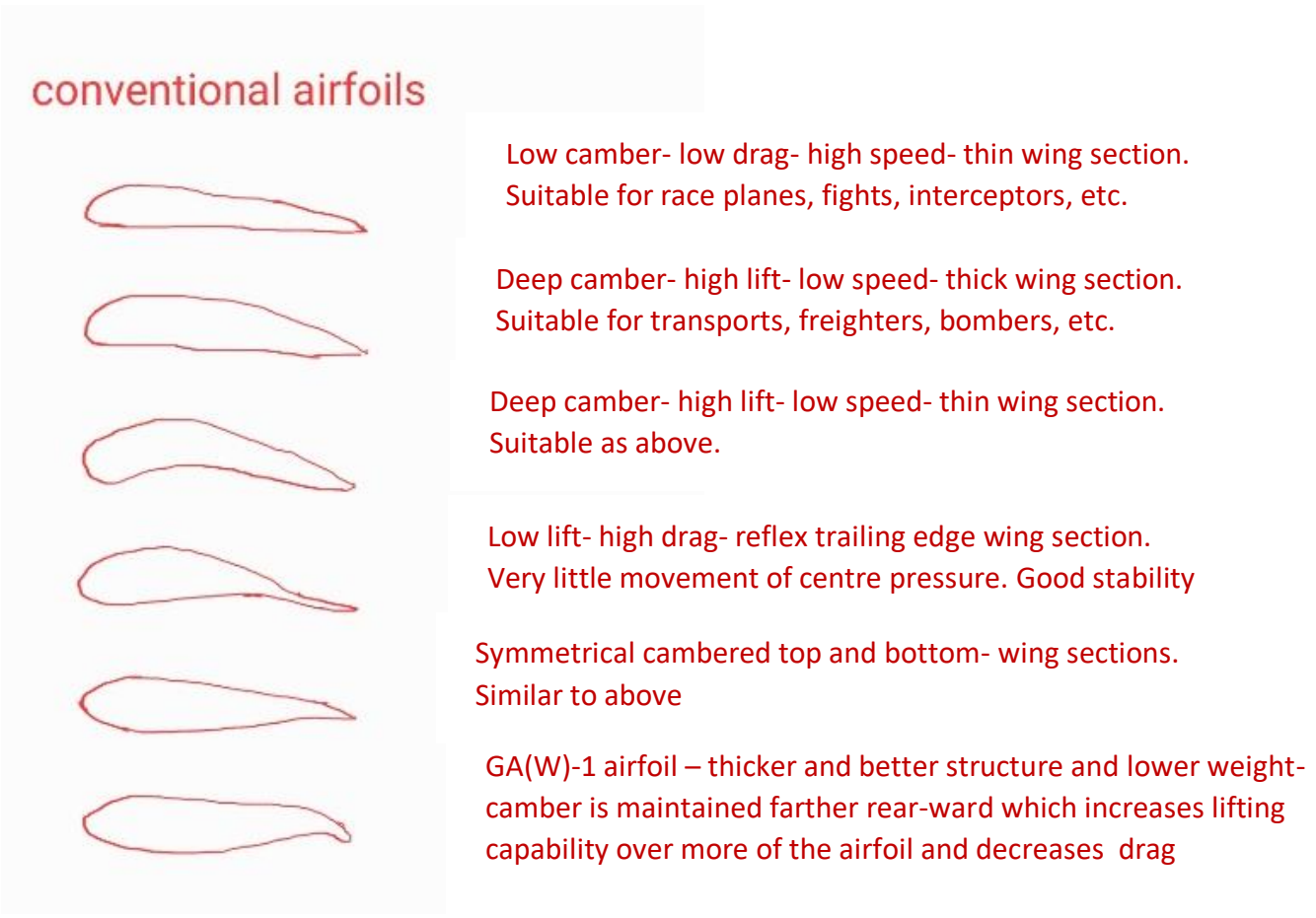

FIGURE 1 Conventional airfoil models [3]

The National Advisory Committee for Aeronautics (NACA) developed eight series of airfoils for aircraft wings.

In Fig. 2A is a scale sketch of a NACA 2412 airfoil whilst Fig. 2B is about the nomenclature of an airfoil. (Please do also consult the NACA website with all the series listed. Please do see also fig. 7. For an exact plot of the Naca 2412 airfoil). The camber line is shown in red, the thickness – or the symmetrical airfoil 0012 surface is shown in purple. The item denoted with 1 is the leading edge and the item denoted with 2 is the trailing edge and the chord line is depicted in black. In the adjacent figure is the angle of attack defined as the angle between the flow motion and chord line (both in aquamarine blue)

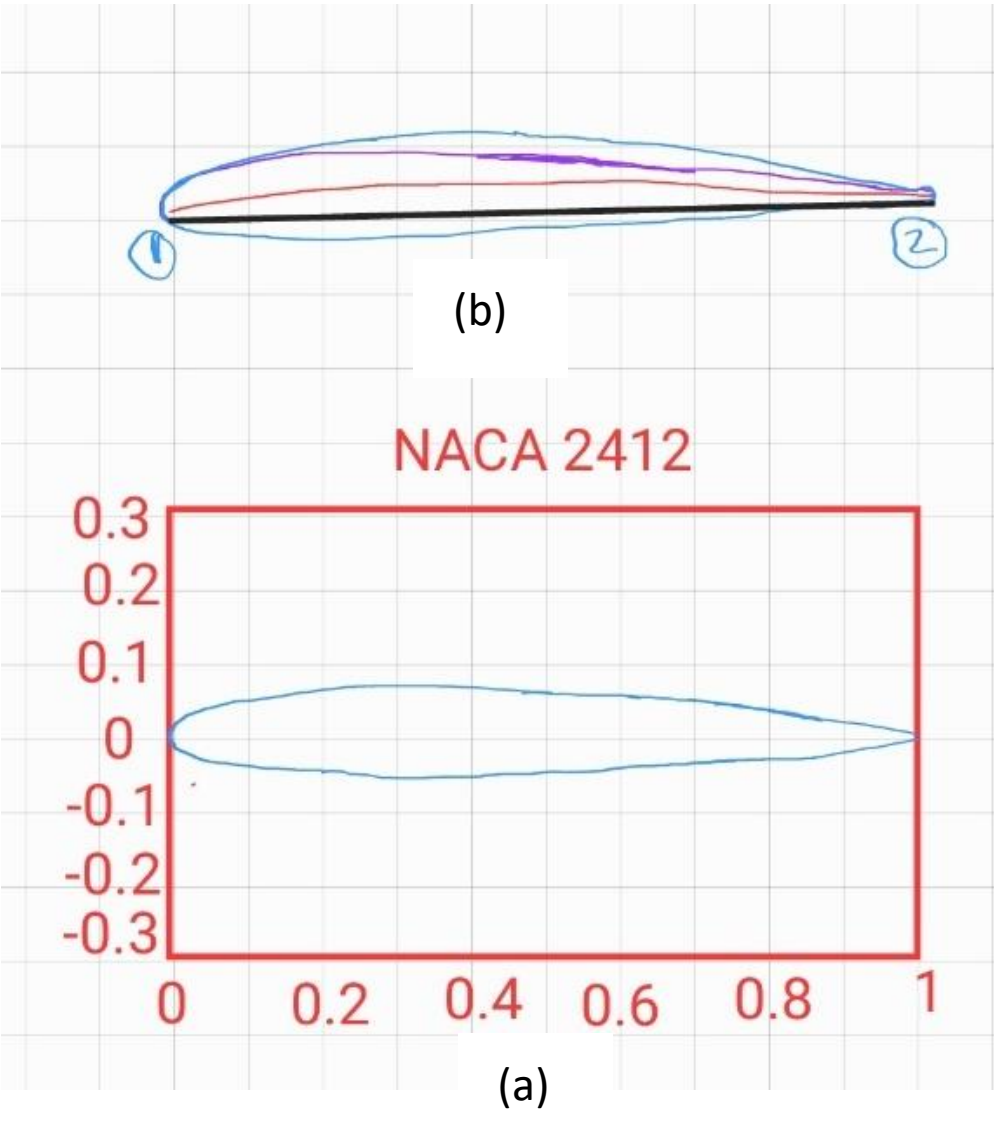

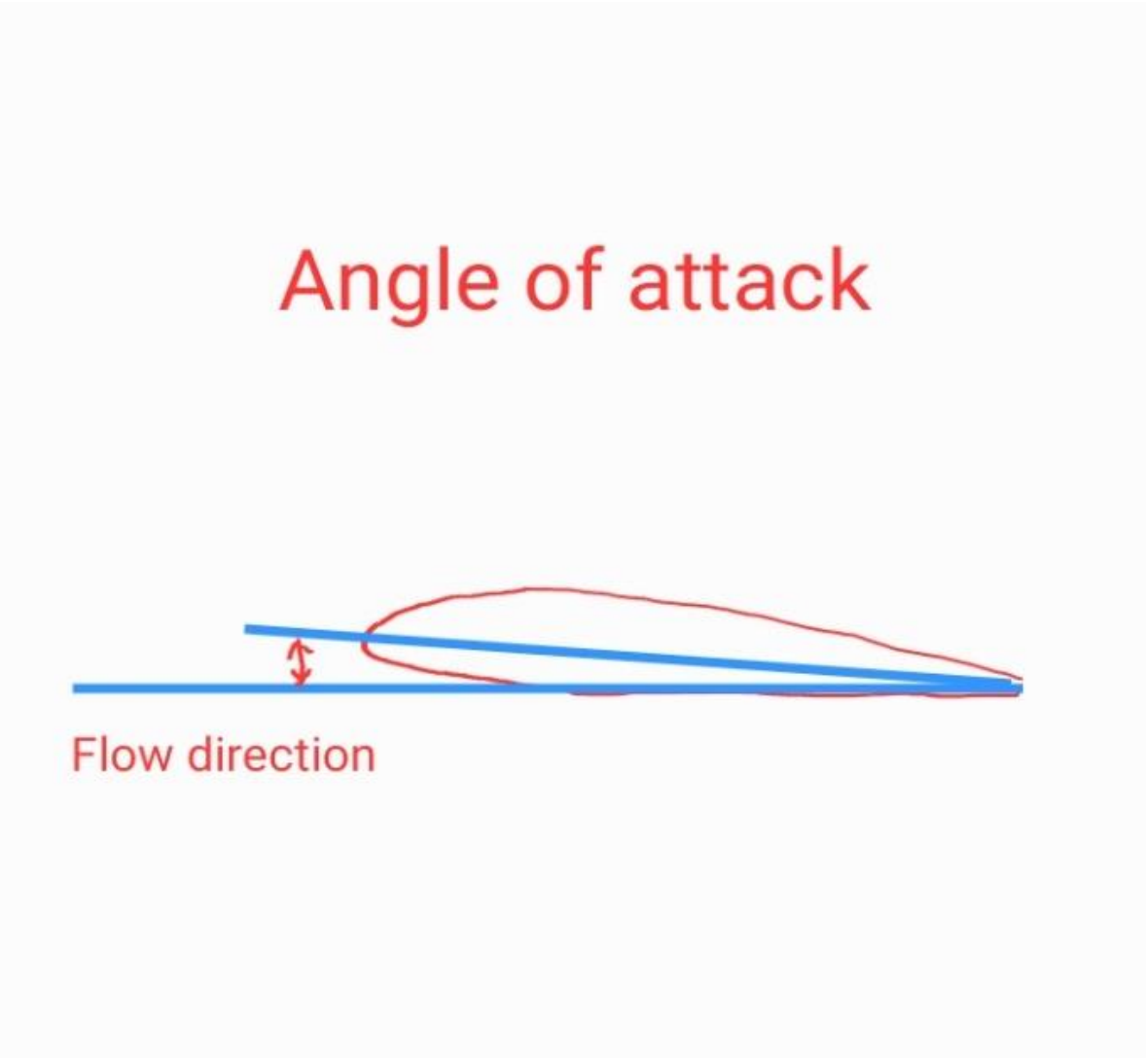

FIGURE 2 (a)-(b) NACA airfoil 2412 and its nomenclature FIGURE 3 Angle of attack definition

References [4] [5]. [6]use different turbulence models to calculate the pressure and shear stresses distribution for the airfoil with type NACA 2412. While ref. [7] used wind tunnel to measure these properties. We will use those results as benchmark.

Figure 14 shows for increasing Reynolds numbers for different (hypothetical) airfoils geometry the value for the statistically measured drag coefficient. Visible is the region before flow separation [8] [9], where the shear stresses decreased abruptly, which is common reason for aircraft dynamic stall and relates also to the condition that the angle of attack approaches to its critical (stall angle) value or the location on the surface where a singularity, due to the geometric definition of the object, resides as we are going to see in the next sections. The Reynolds number for an airfoil is defined as, $R_e = \frac{vc}{\nu}$, with c the chord length, $\nu$ the kinematic viscosity and v the inflow velocity (Please do consult Wikipedia about this subject).

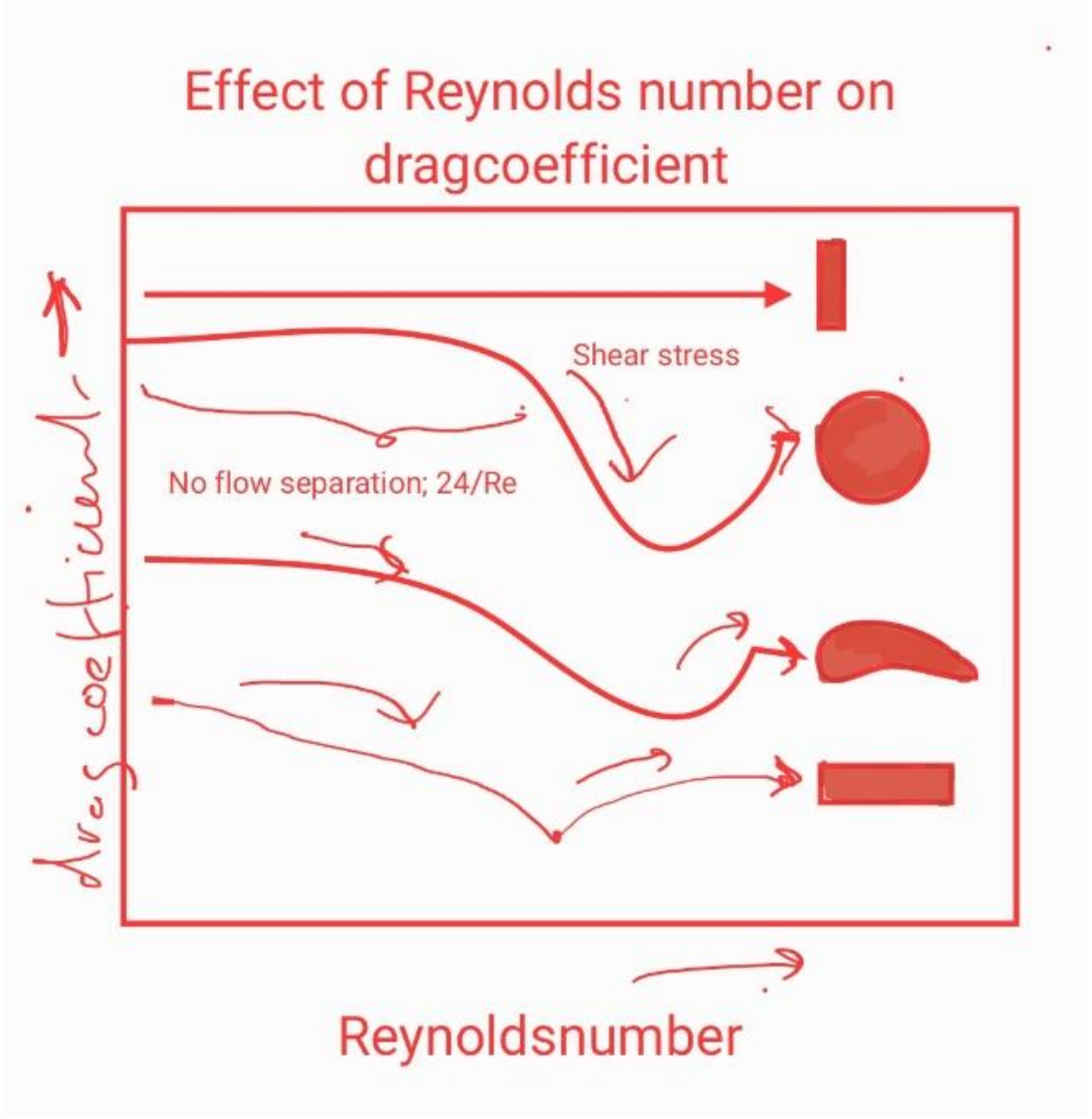

FIGURE 4 Effect of the Reynolds number on the drag coefficient [10]

The pressure stresses $\overrightarrow{\mathrm{p_r}}$ (normal or orthogonal to the surface area) and shear stresses $\vec{\tau}_{\omega}$ (that are tangential to the surface), the lift L (orthogonal on the wind direction) and the draft D (opposite to the wind direction) distributions around an (hypothetical) airfoil is given in Fig. 5. The resultant of the total aerodynamic forces (TAF) is the vectorial sum of the lift and drag forces, while the angle of attack is the angle between the wind direction and the chord.Please do see below fig.5 for an illustration of these forces on a hypothetical airfoil.

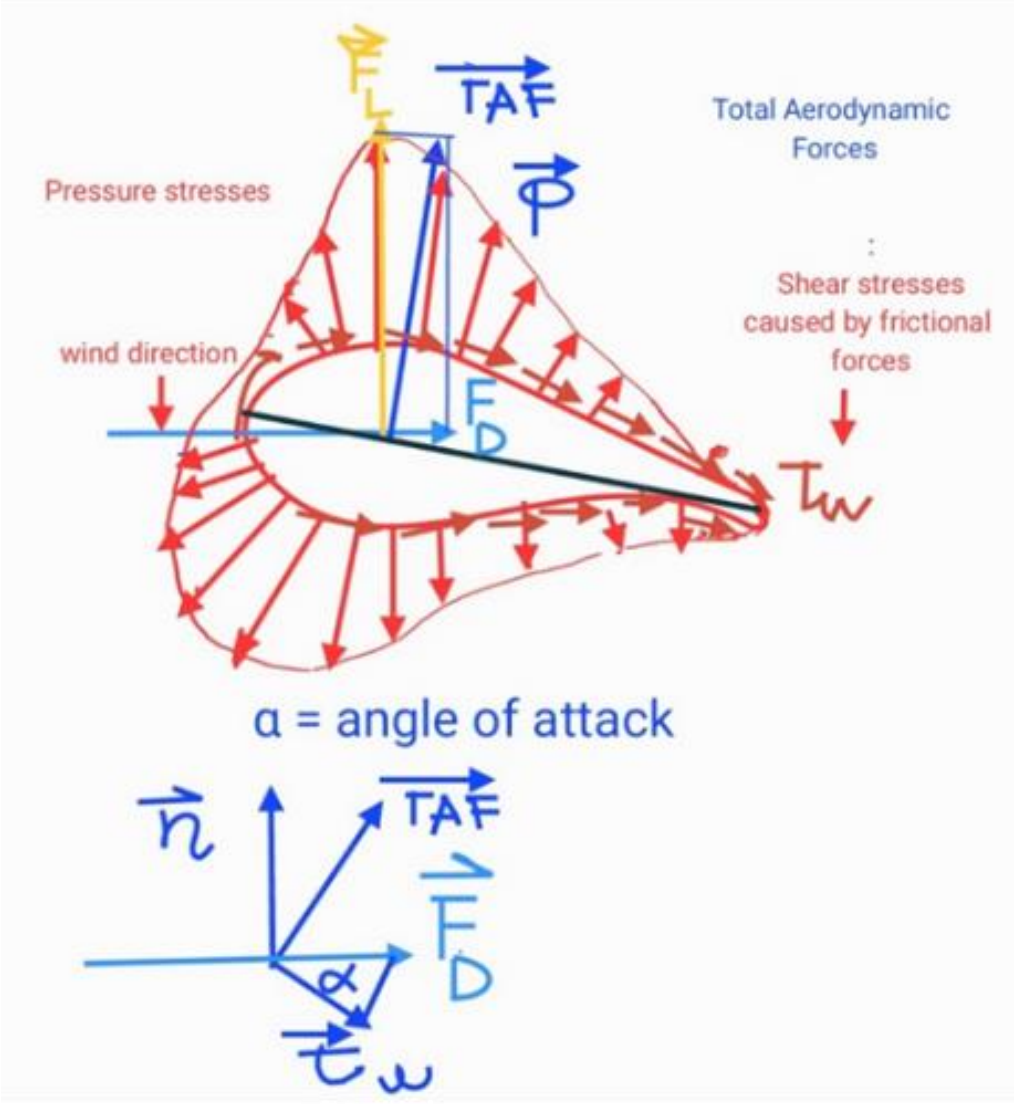

FIGURE 5 Pressure and shear stresses on a hypothetical airfoil [2]

Open ocean saltwater hunting fishes are shaped to avoid flow separation through extension of their hydrodynamical bodies with the high-pressure stress areas in the form of flexible fins. See also below fig. 6.

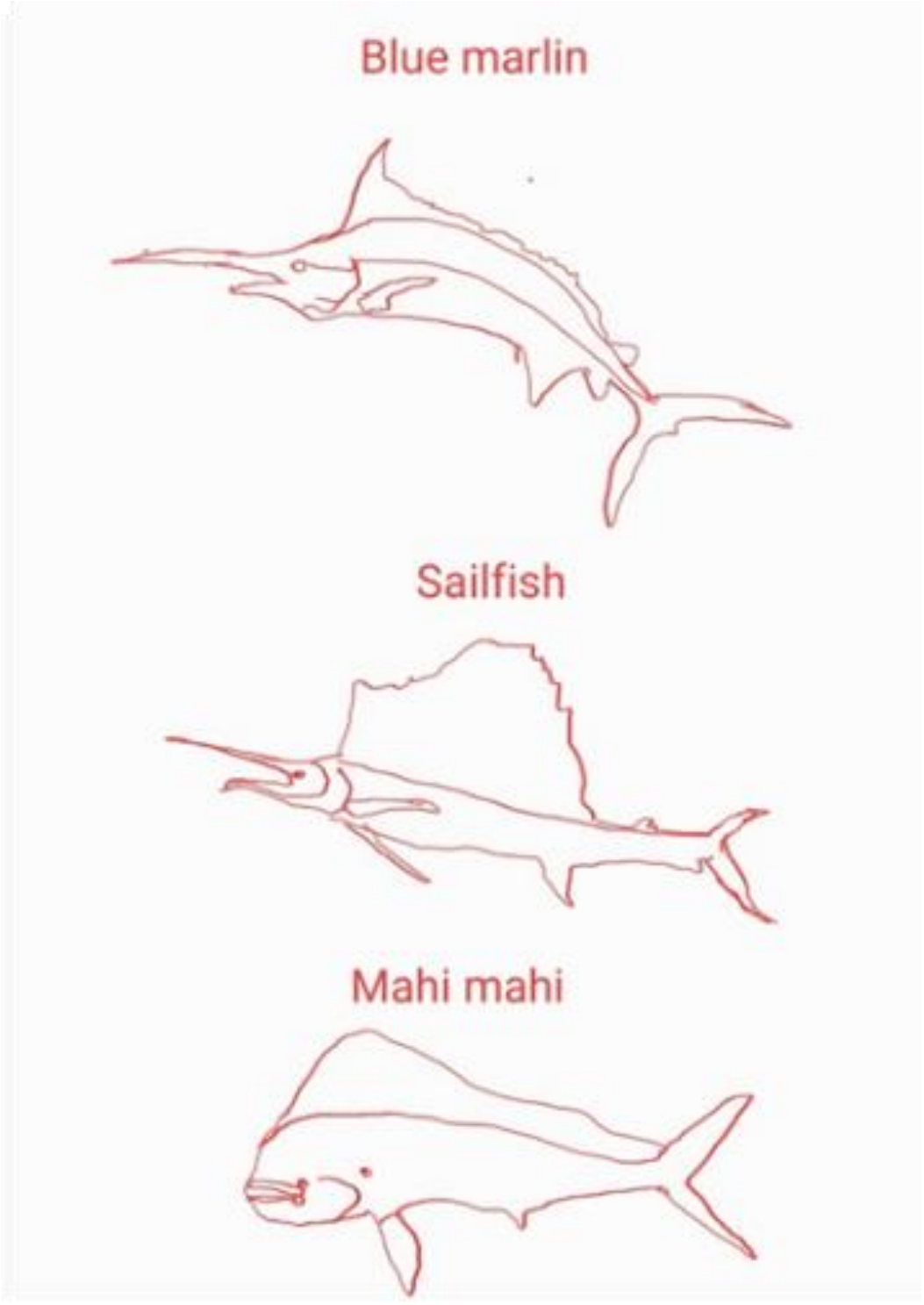

FIGURE 6 Open ocean Saltwater hunting fishes use their super hydrodynamical upper body as extension to cover the low-shear stress areas and high pressure stress areas, to avoid flow separation and to be able to maneuver better. The shape of the hunting fishes does resemble the outcome of our analytical pressure stresses calculations around a NACA 2412 airfoil listed in section III of this manuscript and beyond

The formula for the shape of a NACA 00xx foil, with "xx" being replaced by the percentage of thickness to chord, is (Wikipedia: Naca airfoil)

$$y_t = 5t\left[0.2969\sqrt{x} - 0.1260x - 0.3516x^2 + 0.2843x^3 - 0.1015x^4\right] \qquad (1)$$

where:

*x* is the position along the chord from 0 to 1.00 (0 to 100%),

is the half thickness at a given value of *x* (centerline to surface),

*t* is the maximum thickness as a fraction of the chord (so *t* gives the last two digits in the NACA 4-digit denomination divided by 100).

The simplest asymmetric foils are the NACA 4-digit series foils, which use the same formula as that used to generate the 00xx symmetric foils, but with the line of mean camber bent. The formula used to calculate the mean camber line is

where

$$y_c = \begin{cases} \frac{m}{p^2} * (2 * p * x - x^2), & 0 \leq x \leq p \\ \frac{m}{(1-p)^2} * \left((1 - 2 * p) + 2 * p * x - x^2\right), & p \leq x \leq 1 \end{cases} \tag{2}$$

m is the maximum camber (100 m is the first of the four digits), p is the location of maximum camber (10 p is the second digit in the NACA xxxx description).

For NACA2412 cambered airfoil, because the thickness needs to be applied perpendicular to the camber line, the coordinates ($x_U$,$y_U$) and ($x_L$,$y_L$), of respectively the upper and lower airfoil surface, become

$$\begin{cases} x_U = x - \sin\theta * y_t, & y_U = y_c + \cos\theta * y_t \\ x_L = x + \sin\theta * y_t, & y_L = y_c - \cos\theta * y_t \end{cases} \tag{3}$$

With

$$\theta = \operatorname{atanh} \frac{dy_c}{dx} \tag{4}$$

$$\frac{dy_c}{dx} = \begin{cases} \frac{m}{p^2} * (2 * p * x - x^2), & 0 \leq x \leq p \\ \frac{m}{(1-p)^2} * \left((1 - 2 * p) + 2 * p * x - x^2\right), & p \leq x \leq 1 \end{cases} \tag{5}$$

The potential (combination) which is the input of the Calogero-Moser Hamiltonian many-body system with elliptic particle interaction $\frac{1}{2}\Delta - g\sum_{i \neq j}^{N} \wp(x_i - x_j)$, is calculated applying the lemma 1 out of the main article [11] on the initial conditions of the airfoil. E.g., let us calculate the first section (out of 4, upper left, upper right, lower left and lower right) of the potential valid for $0 \leq x \leq p \wedge p = \frac{4}{10} \wedge m = \frac{2}{100} \wedge t = \frac{12}{100}$, the key figures for the airfoil NACA2412. Please see table 1 for a specific parametrization of the NACA 2412 airfoil necessary to model the boundary conditions for the Navier Stokes d.e. Figure 7 contains a visualization of the calculated surface arcs of the NACA 2412 airfoil. The formulae for the upper and lower surface of this 4th series NACA airfoil model are then uploaded in table 1:

**TABLE 1. THE PARAMETRIZATION OF A NACA ARFOIL 2412**

| Section # $x_U = x - \sin\theta * y_t$ $x_U = x + \sin\theta * y_t$ $y_t = 5t[0.2969\sqrt{x} - 0.1260x - 0.3516x^2 + 0.2843x^3 - 0.1015x^4]$ | $y_c$ | $\frac{\partial y_c}{\partial x}$ | x | y |
|---|---|---|---|---|
| 1 upper left | $\frac{m}{p^2} * (2 * p * x - x^2)$ | $\frac{m}{p^2} * (2 * p - 2 * x)$ | ((1.25*x - 0.5)*0.12*(-0.1015*x^4 + 0.2843*x^3 - 0.3516*x^2 - 0.126*x + 0.2969 sqrt(x)))/sqrt(0.0625*x^2 - 0.05*x + 1.01) + x | x (0.1 - 0.125 x) + (5 0.12(-0.1015 x^4 + 0.2843 x^3 - 0.3516 x^2 - 0.126 x + 0.2969 sqrt(x)))/sqrt(0.0625 x^2 - 0.05 x + 1.01) |
| 2 lower right | $\frac{m}{(1-p)^2} * ((1 - 2 * p) + 2 * p * x - x^2)$ | (2 m (p - x))/(1 - p)^2 | ((0.555556 x - 0.222222) 0.12(-0.1015 x^4 + 0.2843 x^3 - 0.3516 x^2 - 0.126 x + 0.2969 sqrt(x)))/sqrt(0.0123457 x^2 - 0.00987654 x + 1.00198) + x | -(5 0.12(-0.1015 x^4 + 0.2843 x^3 - 0.3516 x^2 - 0.126 x + 0.2969 sqrt(x)))/sqrt(0.0123457 x^2 - 0.00987654 x + 1.00198) - 0.0555556 x^2 + 0.0444444 x + 0.0111111 |
| 3 lower left | See section 1 | See section 1 | 0.5 - 1.25*x)*0.12*(-0.1015*x^4 + 0.2843*x^3 - 0.3516*x^2 - 0.126*x + 0.2969 sqrt(x)))/sqrt(0.0625*x^2 - 0.05*x + 1.01) + x | x (0.1 - 0.125 x) - (5 0.12(-0.1015 x^4 + 0.2843 x^3 - 0.3516 x^2 - 0.126 x + 0.2969 sqrt(x)))/sqrt(0.0625 x^2 - 0.05 x + 1.01) |
| 4 upper right | $\frac{m}{(1-p)^2} * ((1 - 2 * p) + 2 * p * x - x^2)$ | (2 m (p - x))/(1 - p)^2 | ((0.222222 - 0.555556 x) 0.12(-0.1015 x^4 + 0.2843 x^3 - 0.3516 x^2 - 0.126 x + 0.2969 sqrt(x)))/sqrt(0.0123457 x^2 - 0.00987654 x + 1.00198) + x | (5*0.12(-0.1015 x^4 + 0.2843 x^3 - 0.3516 x^2 - 0.126 x + 0.2969 sqrt(x)))/sqrt(0.0123457 x^2 - 0.00987654 x + 1.00198) - 0.0555556 x^2 + 0.0444444 x + 0.0111111 |

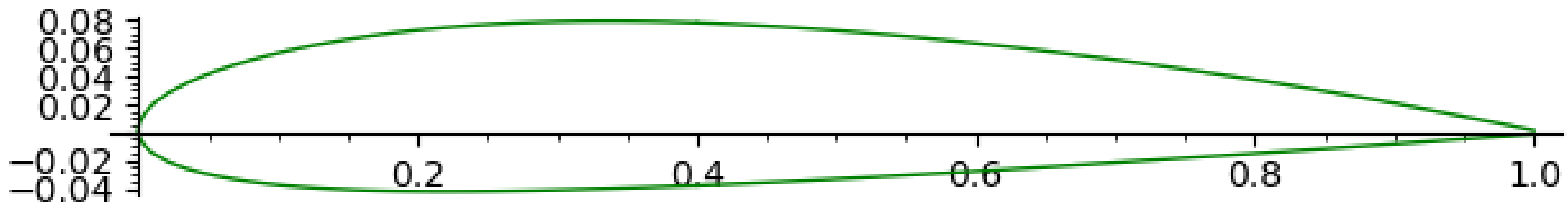

FIGURE 7 The upper and lower surfaces of the airfoil Naca 2412

The initial conditions for the velocity distribution may be incorporated into the general solution (as a natural moving boundary) by using the following elliptic integral equation based on the Lemma 1.

Lemma 1. *The uniformization of curves of genus unity* [12]. If the variables *x* and *y* are connected by an equation of the form $y^2 = a_0x^4 + 4a_1x^3 + 6a_2x^2 + 4a_3x + a_4$ [Eq. (133)], then they can be expressed as one-valued functions of a variable *z* by the equations

$$\begin{cases} x = x_0 + \frac{1}{4}f'(x_0)\frac{1}{\wp(z) - \frac{f''(x_0)}{24}} \\ y = -\frac{1}{4}f'(x_0)\frac{\wp(z)'}{\left\{\wp(z) - \frac{f''(x_0)}{24}\right\}^2} \end{cases} \quad (6)$$

where,

$$f(x) = a_0x^4 + 4a_1x^3 + 6a_2x^2 + 4a_3x + a_4 \quad (7)$$

$x_0$ is any zero of f(x), and the function $\wp(z)$ is formed with the invariants of the quartic:

$$g_2 = a_0a_4 - 4a_1a_3 + 3a_2{}^2 \quad (8)$$

and

$$g_3 = a_0a_2a_4 + 2a_1a_2a_3 - a_2{}^3 - a_0a_3{}^2 - a_1{}^2a_4 \quad (9)$$

The quantity z satisfies

$$z = \int_a^x \{f(x)\}^{-\frac{1}{2}}\, dt \quad (10)$$

The potential as the n-soliton solution to the Navier Stokes d.e. [11] can be written then as

$$\wp(x + \delta) + \wp(x) + \wp(\delta) = (\varsigma(x + \delta) - \zeta(x) - \zeta(\delta))^2 \quad (11)$$

Using Eq. (6) yields the following expression for the required potential hole, using Eq. (10)

$$\wp(z) = \frac{f''(x_0)}{24} + \frac{1}{4}f'(x_0)\frac{1}{x - x_0} \quad (12)$$

## II. BEGIN CONDITIONS AND THE AIRFOIL BOUNDARY LAYER VELOCITY DISTRIBUTION

The boundary layer velocity distribution is a thoroughly and independently studied problem [13] [14]. Some authors and engineers do interpret the boundary layer conditions as no-slip walls built-in in their CFD (Computational Fluid Dynamics) codes.

The boundary layer problem is on itself a Navier Stokes modeling problem of respectable complexity, and no other than the related and main fluid dynamical problem which it forms part of and this can be clarified as follows. The boundary layer is defacto a mass-extension (polyp layer creation) of the airfoil with a lift and drag distribution. The mass-fragmentation process during flow separation (the sudden drop in shear and pressure stresses) may cause instabilities and possibly stall.

Out of lemma 1 of the main article we may deduct the potential field as solution of the underlying KdV flow hierarchy caused by the in the flow the inserted airfoil. Whence the measured layer above and below the airfoil is of the same magnitude and denoted by y, then the integral

$$\int_{y_U}^{y} \frac{1}{\sqrt{(y)^2}}dy - \int_{y_L}^{y} \frac{1}{\sqrt{(y)^2}}dy = l\,n\left(\frac{y_L}{y_U}\right) \quad (13)$$

The r. h. s. of Eq. (13) is then equivelant with

$$2\,\mathrm{atanh}\frac{y_t cos\,\mathrm{atan}\frac{\partial y_c}{\partial x}}{y_c} = 2\,\mathrm{atanh}\frac{\frac{y_t}{\sqrt{1+\left(\frac{\partial y_c}{\partial x}\right)^2}}}{y_c} \quad (14)$$

for $|y| \geq y_U,\ y_L\ \wedge x > 0 \sqcup \{0\}$ using $\mathrm{atanh}\frac{y}{a} = \frac{1}{2}\log\frac{a+y}{a-y}$ and $y_t cos\theta$ resp. $y_C$ the thickness ($y = y_t cos\theta + \underbrace{\tilde{y}}_{\text{measured height}}$) resp. camber line (= a) definition of the airfoil and $\theta = \mathrm{atan}\frac{\partial y_c}{\partial x}$ and $\cos(\mathrm{atan}\, y) = \frac{1}{\sqrt{1+y^2}}$.

The (initial and normalized) velocity distribution in the boundary layer is then

$$u_0(\xi) = \frac{1}{4}(\xi^2 - 1)^2 \text{with } \xi = \frac{\frac{y_t}{\sqrt{1+\left(\frac{\partial y_c}{\partial x}\right)^2}}}{y_c} \text{ and } 0< \xi <1 \tag{15}$$

which is analogous with the begin condition of the benchmark driven-lid cavity problem presented in the main article [11]. Please do see also Fig. 8 illustrating some details.

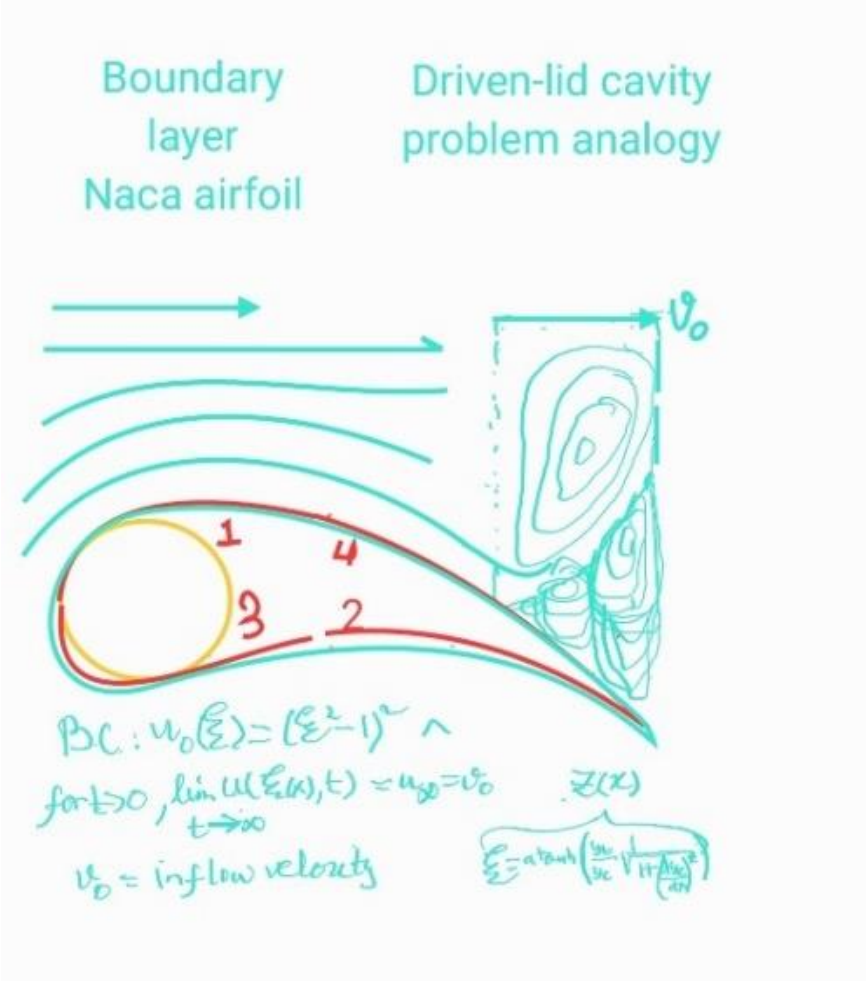

FIGURE 8 Reduction of the boundary layer problem of a NACA airfoil to the driven-lid cavity problem

The z coordinate of the soliton surface z= atanh $(\xi(x))$ is then related to a degenerate case of the Weierstrass p-function as we have seen in the main article with regard to the presented driven-lid cavity problem. The soliton system is "underdetermined" with regard to the p-function (soliton surface) construction using Lemma 1 as a construction aid. The p-coordinate and the z-coordinate of the soliton surface relates parabolically with each other, $p \cong z^2$, and can be seen as follows. (In the referenced article a so-called lump solution or a rational solution is found as a repetitive Cauchy-integral limit of the geometric sum of the collapsed state solutions for the benchmark driven-lid cavity problem using an arbitrary point from the sample domain to construct a non-degenerate p-function. The potential energy generated by particle interaction in this problem is negligible compared to the generated kinetic energy, and the problem is characterized by a degenerate Weierstrass p-function.) The following is a 3-soliton solution for the flow over an airfoil with a degenerate p-function which is a potential choreographing the flows within both the boundary layers and the main stream.

The searched for potential does coincide with a degenerate form of the Weierstrass p-function $\wp(z) \equiv 4\,\text{atanh}^2 \frac{\frac{y_t}{\sqrt{1+\left(\frac{\partial y_c}{\partial x}\right)^2}}}{y_c}$ with

$$\xi(x) = \frac{\frac{y_t}{\sqrt{1+\left(\frac{\partial y_c}{\partial x}\right)^2}}}{y_c} \tag{16}$$

that is both a solution to related the KdV hierarchy flows and congruential with the (inverse) Jacobi elliptic ns function $\text{ns}(\text{u},1) \equiv \text{atanh}\, u$. Hence

$$\wp(u+\omega_3;\ g_2, g_3) = e_3 + (e_1 - e_3)\,\text{ns}^2\left(u\sqrt{e_1 - e_3}, \sqrt{\frac{e_2-e_3}{e_1-e_3}}\right) \tag{17}$$

is then a potential solution to the boundary layer problem with $\wp(\omega_i) = e_i$ for i = 1, 2, 3 and $\omega_i$ i=1,2 are the half periods, $\wp(\omega_3) = -\wp(-\omega_1 - \omega_2) = e_3$ and $u = z\sqrt{e_1 - e_3}$.

A root version of Eq. (17) therefor can serve as the Lax functional α(x) of the related CM Hamiltonian with elliptical interaction with $\alpha(x)\alpha(-x) = -(\wp(x+\delta) + \wp(x) + \wp(\delta))$.
In contrast to the (initial) velocity distribution within the boundary layers that behaves as a fully developed part of a velocity distribution, the (resultant) time evolutive velocity distribution may be calculated with the formulae found in Reference [11]. Hence,

$$\vec{u}(\ \xi(x,y),t) = \frac{u_\infty}{\sqrt{2^{\mathrm{m}}\mathrm{m}!}}\,\mathrm{erf}\ \frac{\left(\sqrt{\kappa}\sqrt{\wp(\xi+\delta)+\wp(\xi)+\wp(\delta)}\right)}{\sqrt{2t}} = \frac{u_\infty}{\sqrt{2^{\mathrm{m}}\mathrm{m}!}}\,\mathrm{erf}\left(\sqrt{2\,\frac{\kappa(\mathrm{atanh}^2(\xi+\delta)+\mathrm{atanh}^2(\xi)+\mathrm{atanh}^2(\delta))}{t}}\right) \tag{18}$$

for t > 0 with $\kappa \in \mathbb{C}$ [please do see Eqs. (22) & (23) for the calculation of κ], $\wp(\xi) = \mathrm{atanh}^2(\xi)$ and $(e_1, e_2, e_3) = (1,1,0)$ $\wedge\, \omega_1,\ \omega_2 = \frac{\left(\mp\frac{1}{8}+\frac{i}{8}\right)\Gamma^2\left(\frac{1}{4}\right)}{4\sqrt{\pi}}\left(\text{using the Wolfram.com WeierstrassHalfPeriods}\left[\left\{\overbrace{-4}^{g_2}, \overbrace{0}^{g_3}\right\}\right]\right)$, correcting the sign and $\xi(x) = \frac{y_t}{y_c\sqrt{1+\left(\frac{\partial y_c}{\partial x}\right)^2}}$ and

$$\lim_{t\to 0}\vec{u}(\ \xi(x,y),t) = \frac{u_\infty}{\sqrt{2^{\mathrm{m}}\mathrm{m}!}}\,\mathrm{erf}\ \frac{\left(\sqrt{\kappa}\sqrt{\wp(\xi+\delta)+\wp(\xi)+\wp(\delta)}\right)}{\sqrt{2t}}$$
$$= \frac{u_\infty}{\sqrt{2^{\mathrm{m}}\mathrm{m}!}}\,\mathrm{erf}\left(\sqrt{2\,\frac{\kappa(\mathrm{atanh}^2(\xi+\delta)+\mathrm{atanh}^2(\xi)+\mathrm{atanh}^2(\delta))}{t}}\right) = \frac{\hat{u}_\infty}{\sqrt{2^{\mathrm{m}-1}\mathrm{m}!}} = v_0,$$
$$m = 1,2,3,..$$

The elliptic invariants $g_2$ and $g_3$ are calculated with the following formulae

$$g_2 = -4(e_1e_2 + e_1e_3 + e_2e_3) \tag{19}$$
$$g_3 = e_1e_2e_3 \tag{20}$$

The identified case of the potential is then analogous with that of the pseudo lemniscate case $\left\{\overbrace{-1}^{g_2}, \overbrace{0}^{g_3}\right\}$ with $\omega_1$ resp. $\omega_2$ equivalent to $\pm 1 + \frac{L}{4}i$ and L the Lemniscate constant equal to 1.311028 (OEIS A085565) (Out of Wolfram.com). (The implication that $e_3 = \wp(\omega_3) = 0$ may be supported by the fact that the p-function does have complex roots. [15])

Eq. (18) does represent the tangential part of the velocity distribution, along the streamlines. The surfaces (upper surface of a symmetric airfoil is used to deduct the Fundamental Thin Airfoil equation [16] [17] ) are interpreted as a stream-line extension, and are used in section IV of this manuscript to calculate amongst others the lift and drag forces on the airfoil with regard to a pitching angle of choice.

Outside the boundary layers the motion of poles (the center of the swirls), which commence within the boundary layers, have to be applied to reconstruct and acquiring a total picture for the velocity distribution and the streamlines for t >0, presumedly occurring after the flow separation. Please do see also the Fig. 9 and Fig. 10 for a sketch visualization resp. a calculation of the trajectory of the poles [11]

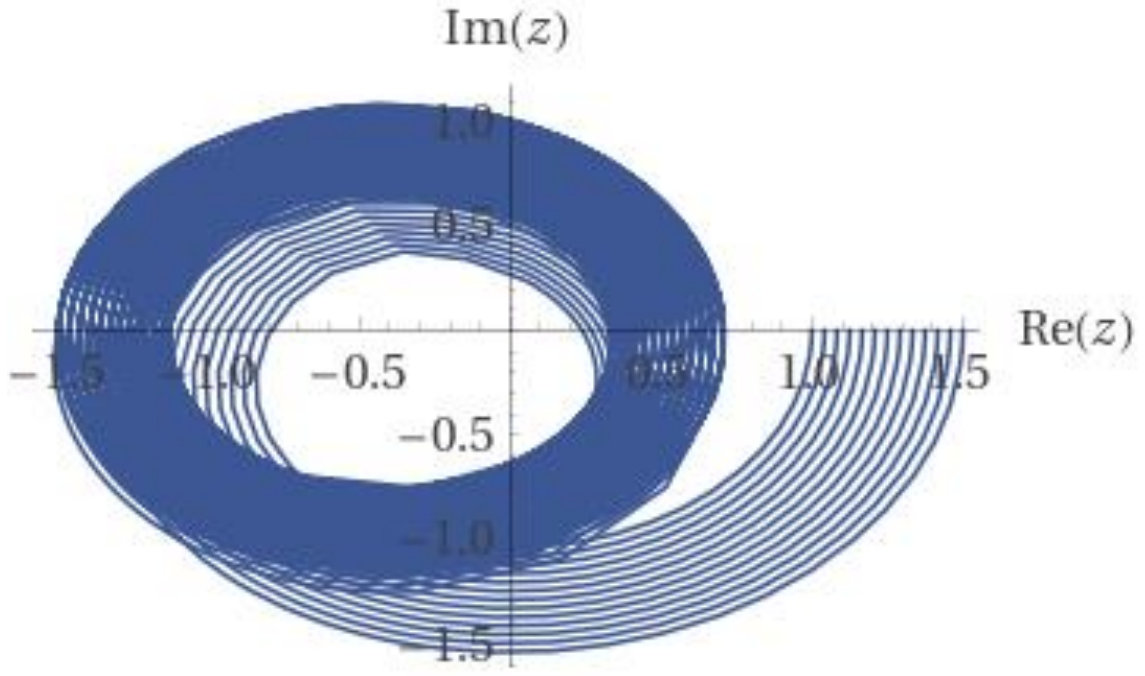

$$z(x) = c_1\, e^{-4ix} - \frac{1}{8}\, i\, e^{8i-4i(x+1)}\, \mathrm{Ei}(4i(x-1)) + \frac{1}{8}\, i\, e^{-4i(x+1)}\, \mathrm{Ei}(4i(x+1)) - \frac{1}{4}\, i \tanh^{-1}(x)$$

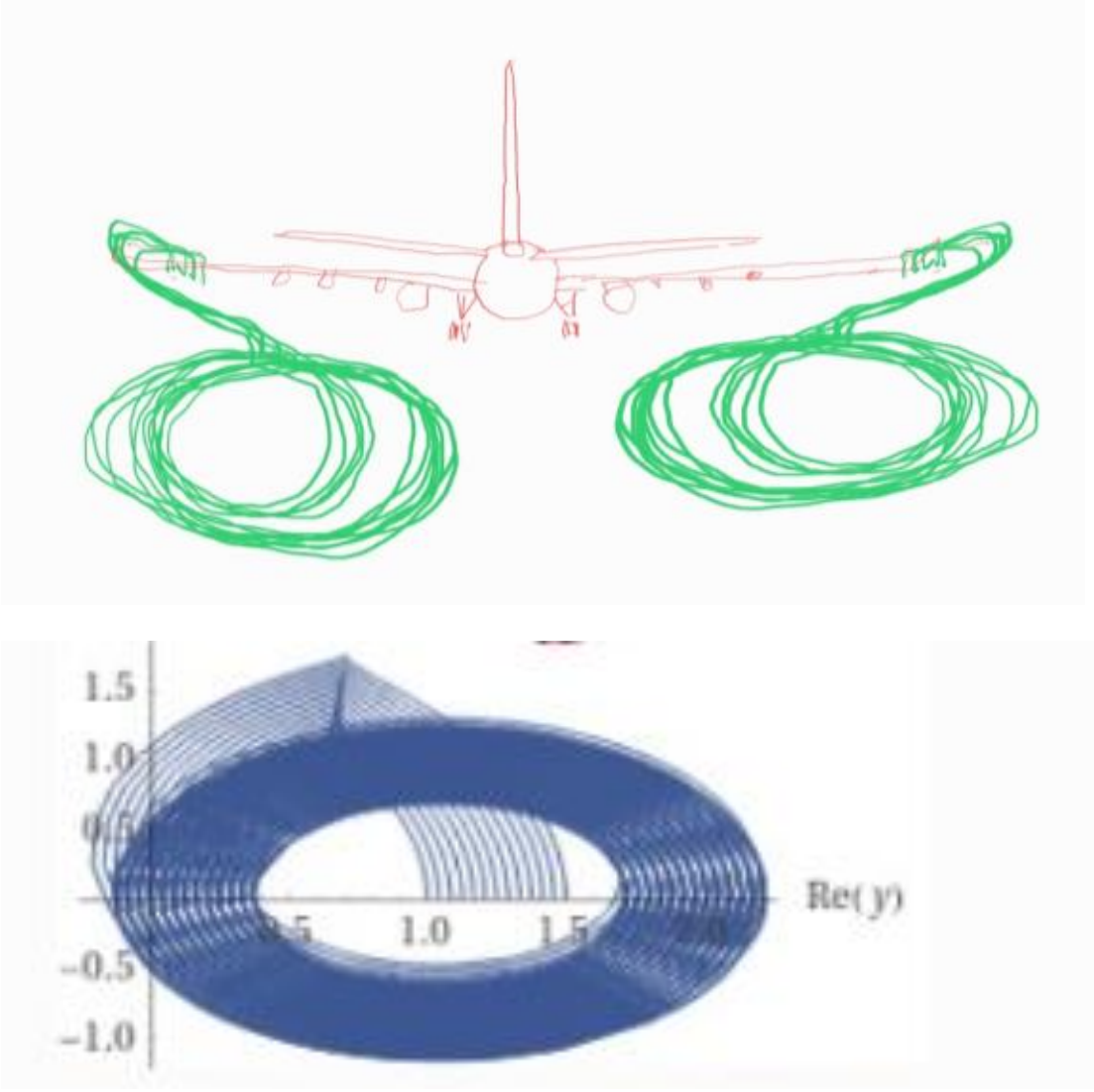

FIGURE 9 RESP.FIGURE 10 Motion of poles of the related CM Hamiltonian in the complex plane. (Initialization point of) the Streaklines of the NACA 2412 airfoil yielded by solving z'= atanh(x)+ φ i*z [11] for $\varphi = -4$ resp. $\varphi = \frac{3}{2}$. Above in fig. 10 the rearview of the wingtips of an aeroplane and the space distortion that is caused by the motion the poles of the related CM Hamiltonian.

The parameter $\delta$ maybe related to both (1) kinematic viscosity through the continuity equation imposed on the flow within the interior of the sample area Ω and (2) to the elliptic invariant $g_2$ as a begin condition constant within the solution of the related KdV flow hierarchy Eq. (22) as follows

$$c_1 = -\mu t + \delta \tag{21}$$

$$\underbrace{\overbrace{{}_n\vec{x}.\nabla\, {}_n\vec{x} = -\underbrace{\frac{1}{\rho}\frac{m}{A}}_{\frac{L}{R_e}}\mu\, \nabla^3\, {}_n\vec{x}}^{\text{stationary KdV}}} \wedge {}_n\vec{x} = \frac{2^{\frac{2}{3}}3^{\frac{1}{3}}\,\wp\left(\left(\left(\frac{\left(-\frac{\overbrace{-\frac{R_e}{L}}}{\frac{A\rho}{m\mu}}\right)^{\frac{1}{3}}}{2^{\frac{2}{3}}3^{\frac{1}{3}}}(x+c_1)\right); -\frac{\overbrace{((2.2^{\frac{2}{3}})\,3^{\frac{1}{3}}c_1)}^{g_2}}{\left(-\frac{\overbrace{\frac{R_e}{L}}}{\frac{A\rho}{m\mu}}\right)^{\frac{1}{3}}}, c_2\right)\right)}{\left(-\frac{\overbrace{\frac{R_e}{L}}}{\frac{A\rho}{m\mu}}\right)^{\frac{1}{3}}}\ [11] \tag{22}$$

$${}_n\vec{x} = -\frac{12L}{R_e}\lambda^2\wp\left(\left(\left(\overbrace{\frac{\left(-\frac{R_e}{L}\right)^{\frac{1}{3}}}{2^{\frac{2}{3}}3^{\frac{1}{3}}}}^{\lambda}(x+c_1)\right); -\overbrace{\frac{\left(\left(2.2^{\frac{2}{3}}\right)3^{\frac{1}{3}}c_1\right)}{\left(-\frac{R_e}{L}\right)^{\frac{1}{3}}}}^{\frac{g_2}{\lambda^4}}, \overbrace{c_2}^{\frac{g_3}{\lambda^6}}\right)\right) = -\overbrace{\frac{12L}{R_e}}^{\kappa=\lambda^{-3}}\wp\left(\left(((x+c_1)); g_2, g_3\right)\right) \tag{23}$$

Using $\wp(z; g_2, g_3) = \lambda^2\wp\left(\left((\lambda(x+c_1)); -\frac{g_2}{\lambda^4}, \frac{g_3}{\lambda^6}\right)\right)$ out of Reference [18] and $R_e$, resp. L the Reynoldsnumber resp. a length scale within the dimensionless version of the Navier Stokes d.e., with $c_1 < 0$.

Combining Eq. (21) and Eq. (18) wil yield a hyperbolic relationship between the kinematic viscosity and the time-like variable

$$\begin{cases} \frac{g_2}{\lambda^4} = -\frac{\left(\left(2.2^{\frac{2}{3}}\right)3^{\frac{1}{3}}c_1\right)}{\left(-\frac{R_e}{L}\right)^{\frac{1}{3}}} \Rightarrow 2.\,c_1 = -\frac{g_2}{\lambda^3} \Rightarrow \frac{c_1}{L} = \frac{24}{R_e} & (23.1.1) \\ c_1 = -\mu t + \delta & (23.1.2) \\ \mu = -\frac{A\,\delta\,\rho}{(24\text{ m} + A\,t\,\rho)} & (23.1.3) \end{cases}$$

In the original paper of Boussinesq (1877) he hypothesized that turbulent fluctuations have a dissipative effect on the mean flow. A principle (the so-called eddy viscosity hypothesis) [19] that is widely used by many mayor CFD turbulence model application software to address the closure problem issues. However, Eq. (23.1.3) does provide the possible necessary conditions for the validity of that assumption.

Whence the boundary layers below and above the airfoil measuring heights differs from each other (that means whence the upper and lower boundary layers are measured asymmetrically) the following formulae are applicable for the (z-coordinate of the soliton) boundary layer at the upper resp. lower Riemann surfaces (please see also Fig. 11 and Fig. 12 for an impression of the Riemann surfaces):

$$\frac{1}{2}\log\frac{x-y}{x-y-a} = 2\,\text{atanh}\frac{\frac{x-y}{x-y-a}-1}{\frac{x-y}{x-y-a}+1} + 2\pi i \qquad (24)$$

with x does represent the camber line, y the thickness and a the measuring height.

$$\frac{\vec{u}_{U,L}(\,\xi(x,y)-2\pi i,t)}{v_0} = \frac{1}{\sqrt{2^m m!}}\,\text{erf}\left(\sqrt{\kappa}\sqrt{2\frac{(\text{atanh}^2(\xi+\delta)+\text{atanh}^2(\xi)+\text{atanh}^2(\delta))}{t}}\right) \qquad (25)$$

with

$$\xi(x,y) = \frac{y}{\frac{2y_t}{\sqrt{1+\left(\frac{\partial y_c}{\partial x}\right)^2}}+2y_c+y} \qquad (26)$$

for the upper surface resp.

$$\xi(x,y) = \frac{y-2y_c+2\frac{y_t}{\sqrt{1+\left(\frac{\partial y_c}{\partial x}\right)^2}}}{y} \qquad (27)$$

for lower surface with

$$\frac{1}{2}\log\frac{x+y+a}{x+y} = 2\,\text{atanh}\frac{\frac{x+y+a}{x+y}-1}{\frac{x+y+a}{x+y}+1} + 2\pi i \qquad (28)$$

with x does represent the camber line, y the thickness and *a* the measuring height.

The z coordinate z(x,y) of the soliton-surface is then a Riemann surface

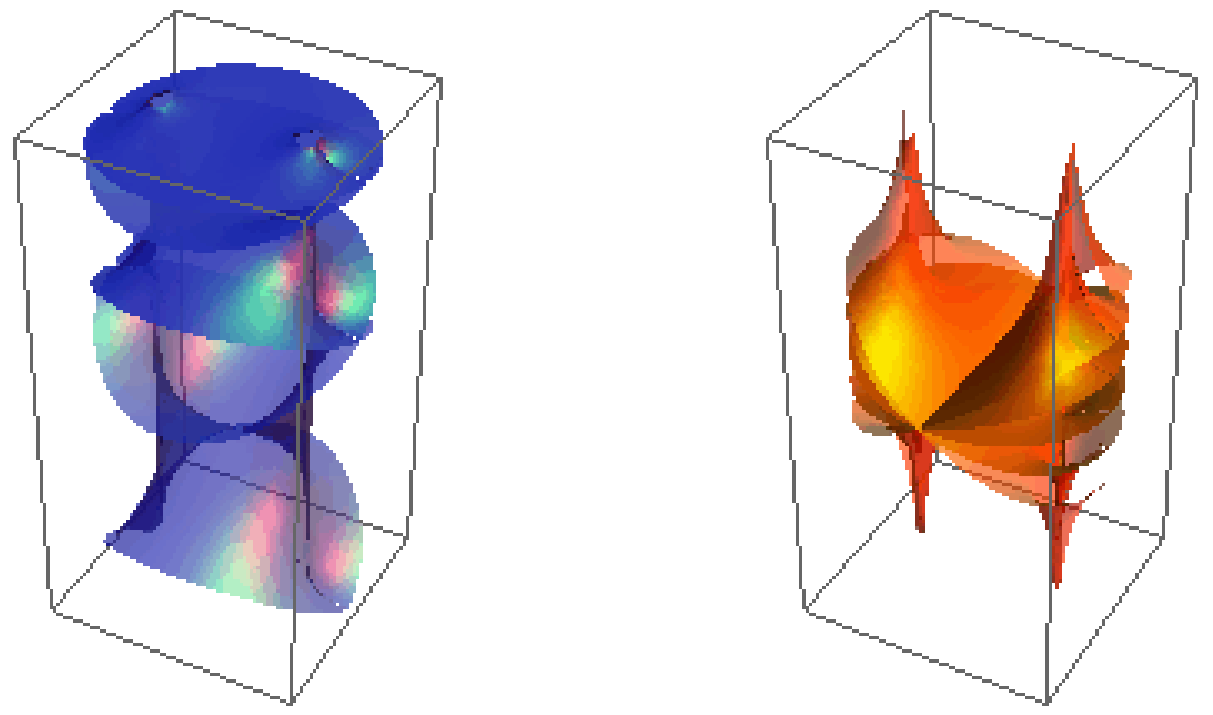

Figure 11 resp.12 Riemann surfaces relation to the z-coordinate of the soliton surfaces

The preliminary results, how the shear and pressure stresses will look alike with simplified soliton pole structure for the NACA airfoil 2412, are shown in Fig. 13 resp. Fig. 14.

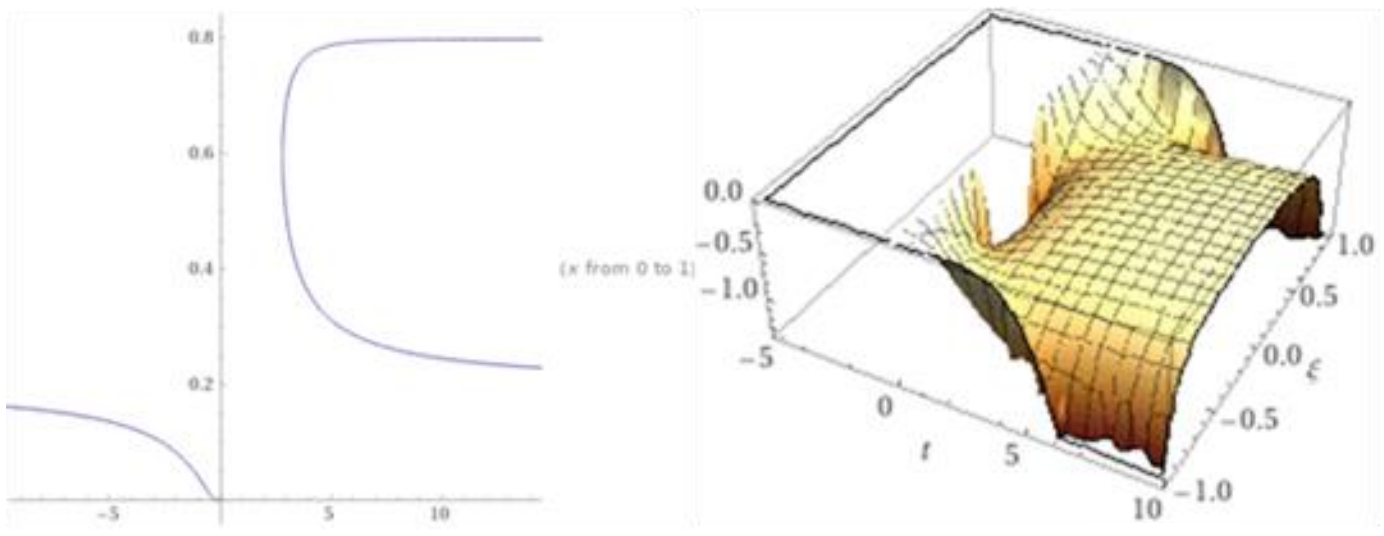

(a) (b)

FIGURE 13 (a) Resultant 1-soliton based shear stress on section 1 & 3 as a function of x and (b) a 3d impression, $\boldsymbol{\nu}$ is set to 1, without the factor $\frac{\mathbf{1}}{\sqrt{\mathbf{2^m m!}}}$

For Newtonian flows the shear stress (tensor) $\vec{\tau}_\omega$ is defined as $\vec{\tau}_\omega = \nu \frac{\partial \vec{u}}{\partial \xi} = (2 * \nu * \exp(-(2 * \text{atanh}(\xi)^2)/t) * \text{sqrt}(2/\text{pi}) * \text{sqrt}(1/t))/(-1 + \xi^2)$ and $\xi = \frac{y_t}{y_c\sqrt{1+\left(\frac{\partial y_c}{\partial x}\right)^2}}$ with $\nu$ the dynamic viscosity

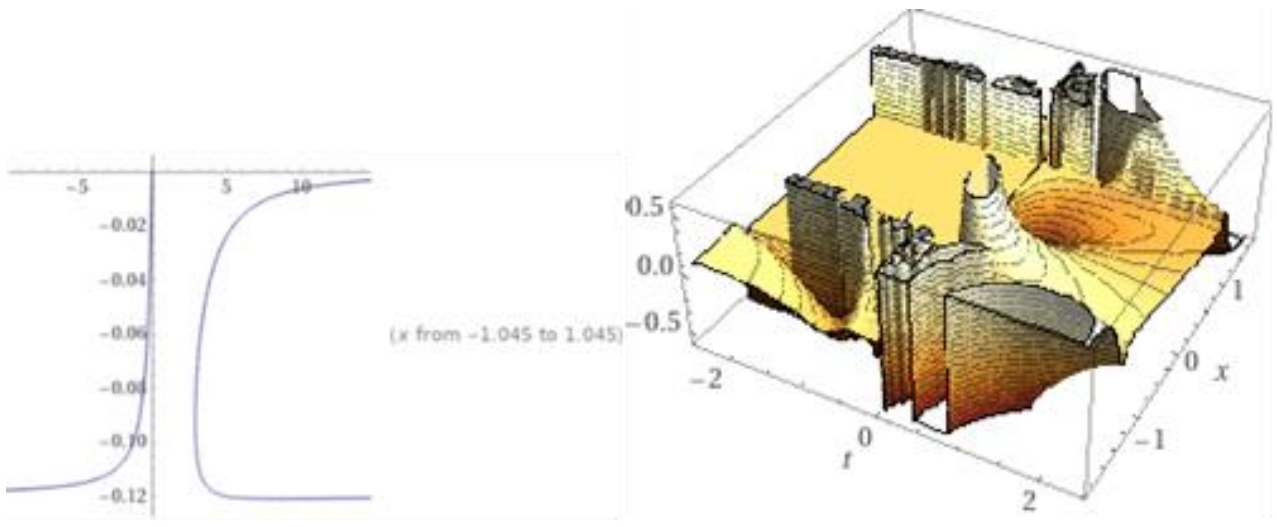

(a) (b)

FIGURE 14 (a) Resultant 1-soliton based pressure stresses for section 1 & 3 of NACA airfoil 2412 as function of x and (b) a 3d impression, without the factor $\frac{\mathbf{1}}{\sqrt{\mathbf{2^m m!}}}$

$$\overrightarrow{\mathrm{p_r}} = \frac{m}{A}\frac{\partial \vec{u}}{\partial t} = -\frac{m}{A}\,(\mathrm{e}\hat{\ }(-\tanh\hat{\ }(-1)(\xi)\hat{\ }2/\mathrm{t})\ \tanh\hat{\ }(-1)(\xi))/(\mathrm{sqrt}(\pi)\ \mathrm{t}\hat{\ }(3/2))\ \text{as function of}\ \xi$$

## III. THE FOUND (RELATIVE) PRESSURE AND SHEAR STRESSES ON THE NACA AIRFOIL 2412

As a consequence of exact solving of the (non-dimensional) Navier Stokes d.e. we are able to calculate explicitly the relative pressure and resp. shear stresses on an airfoil. As usual when solving the NVS d.e., to retrieve the absolute pressure, the atmospheric pressure has to be added [not to confuse with the length of the pressure value, which is also denote with the word “absolute”]:

$$\frac{\mathrm{Du^*}}{\mathrm{Dt}} + \mathrm{u^*}.\nabla \mathrm{u^*} = -\frac{1}{\rho}\nabla \mathrm{p_r}^* + \frac{1}{R_e}\nabla^2 \mathrm{u^*} + \mathrm{F}, \qquad (29)$$

with $u^* \coloneqq \frac{u}{v_0}$ (29.1), $\mathrm{p_r}^* \coloneqq \frac{\mathrm{p_r}}{{v_0}^2}$ (29.2), $\mathrm{R_e} = \frac{v_0 c}{\nu}$ and the length scale $r^* \coloneqq \frac{r}{L}$ (and so the time-scale) , $\nabla \coloneqq L\,\nabla$ (29.3) with $\nu$ being the dynamic viscosity of the medium, c=1 the chord length and $v_0$ the inflow velocity.

Please do see the figure series 16-27 for the results for the Naca airfoil 2412 at an angle of attack of 0°radians with

m= mass of the airfoil

A=area of contact

$\overrightarrow{\mathrm{p_r}}^*$ = pressure stress

$\vec{\tau}_\omega$ = shear stresses

with $\overrightarrow{\mathrm{p_r}}^* = \frac{m}{A}\frac{\partial \overrightarrow{u^*}}{\partial t}$ and $\vec{\tau}_\omega = \nu \frac{\partial \overrightarrow{u^*}}{\partial x}$, the collapsed state velocity vector along the infinitely many stream-lines

${}_{n+1}\overrightarrow{u^*} = \frac{d}{dt}\frac{1}{\sqrt{2^{\mathrm{n}}\mathrm{n}!}}\frac{\partial^{n+1}}{\partial z^{n+1}}\mathrm{erf}\,\frac{z(\xi)-z_0}{\sqrt{2t}}$ (29.4), for $\mathrm{n} = 1,2,\ldots$ [11]. Assuming the lamb vector is null and the $\mu =$ kinmatic viscosity , $\nu$ the dynamic viscosity and $\xi(\mathrm{x})$ is the hyperbolic tangent of the z-coordinate of the potential that is determined by the begin position which is a quartic. The tangential velocity distribution (along the stream-lines for the collapsed states with even parity) is then represented as

$$\overrightarrow{u^*}_{tangential}(x,t) = \frac{1}{\sqrt{2^m m!}} \operatorname{erf} \frac{\sqrt{\kappa(\wp(z(x)+\mu t+\delta)+\wp(z(x)+\mu t)+\wp(\delta))}}{\sqrt{2t}} \qquad (30)$$

with $\kappa \in \mathbb{C}$ a begin condition constant and m=1,2,... or as one of its infinitely many space-like derivatives (second and beyond) which are Hermite functions and in the complex plane are these Euler-Cornu spirals. Below Fig. 15 is an illustration of the derivatives of the Hermite functions. The outer left element is the second derivative, the third one is the fourth derivative and the fifth one (outer right) is the sixth derivative of the error function in the complex plane. The parity of the solutions is depended on the begin conditions, a fact that is forthcoming both in nature in fluctus cloud formations, Von Karman vortex streets etc. and our analytical results. [11] The soliton-solutions are quantized, and tend to conserve the parity of the initial begin condition.

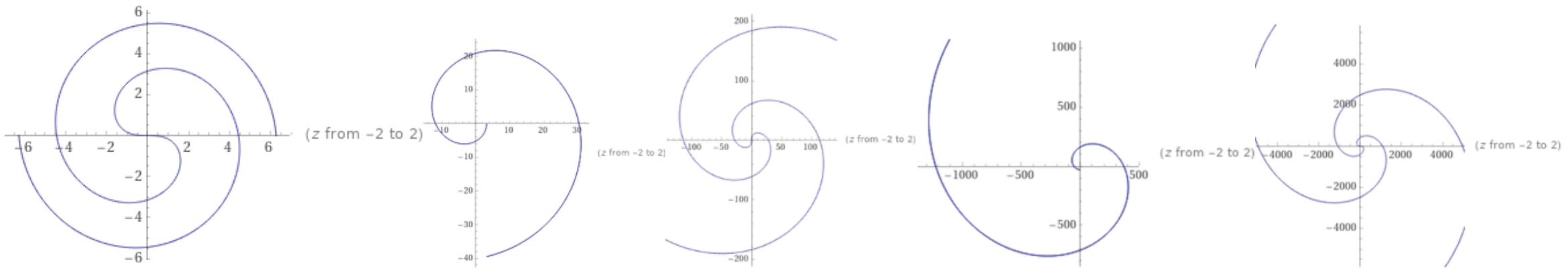

FIGURE 15 Euler Cornu spirals represented as derivatives of the Hermite functions in the complex plane as the collapsed state solutions of the NVS d.e.

For further study we will take Eq. (30) as the tangential velocity of the particles within the hydrodynamical medium.

Using the identity $\frac{A\,\rho}{m\,\mu} \equiv \frac{R_e}{L}$ out of Eq. (22) follows that $\frac{m}{A} \equiv \frac{L}{\nu R_e}$ (30.1). Combining the last identity with the identity $\overrightarrow{p_r}^* = \frac{m}{A}\frac{\partial \overrightarrow{u^*}}{\partial t}$ and Eqs. (29- 29.3) follows that

$$\overrightarrow{p_r}^* = \frac{m}{A}\frac{\partial \overrightarrow{u^*}}{\partial t} = \frac{L}{\nu R_e}\frac{\partial \overrightarrow{u^*}}{\partial t} = \frac{L}{\nu\frac{v_0 c}{\nu}}\frac{\partial \overrightarrow{u^*}}{\partial t} \Rightarrow \overrightarrow{p_r} = v_0^2\frac{L}{v_0 c}\frac{\partial \overrightarrow{u^*}}{\partial t} = v_0\frac{L}{c}\frac{\partial \overrightarrow{u^*}}{\partial t} \qquad (30.2)$$

Using the found potential distributions, we deduced the following pressure distributions around the airfoil, with the parameter $\mu$ =1.516*10^(-5) $m^2/s$ at 20°C, $\rho = 1.202\ kg/m^3$ and $\nu$ =1.825*10^(-5) kg/ms at 20°C for dry air and n=1.

Using

$$\overrightarrow{p_r}^* = \frac{m}{A}\frac{d}{dt}\frac{1}{\sqrt{2^n n!}} \operatorname{erf} \frac{\sqrt{\kappa(\wp(z(x)+\mu t+\delta)+\wp(z(x)+\mu t)+\wp(\delta))}}{\sqrt{2t}} \qquad (31)$$

for the found 3-soliton solution system for the NVS d.e. and $\kappa = -\frac{12L}{R_e}$, and $c_1 < 0$ ($c_1$ *as in* Eq. (23))

$$\overrightarrow{p_r}^* = \frac{m}{A}\frac{1}{\sqrt{2^n n!}}\frac{d}{dt}\mathrm{erf}\,\frac{\sqrt{\kappa\big(\wp(z(x)+\mu t+\delta)+\wp(z(x)+\mu t)+\wp(\delta)\big)}}{\sqrt{2t}} =$$

$$= -\frac{1}{\sqrt{2^n n!}}\frac{m}{A}\frac{\sqrt{\frac{2}{\pi}}e^{-\frac{\left(\sqrt{\kappa(\wp(z(x)+\mu t+\delta)+\wp(z(x)+\mu t)+\wp(\delta))}\right)^2}{2t}}}{\sqrt{2t}}\left[\frac{\kappa\overbrace{\big(\wp(z(x)+\mu t+\delta)+\wp(z(x)+\mu t)\big)'}^{(\varsigma(z(x)+\mu t+\delta)+\zeta(z(x)+\mu t))\left(\frac{dzdx}{dxdt}+\mu\right)}}{2\sqrt{2t}\sqrt{\kappa\big(\wp(z(x)+\mu t+\delta)+\wp(z(x)+\mu t)+\wp(\delta)\big)}} - \frac{\kappa\big(\wp(z(x)+\mu t+\delta)+\wp(z(x)+\mu t)+\wp(\delta)\big)}{2\sqrt{2}\sqrt{t^3}}\right]$$

$$= -\frac{1}{\sqrt{2^n n!}}\rho L\frac{\sqrt{\frac{2}{\pi}}e^{-\frac{\left(\sqrt{\kappa(\wp(z(x)+\mu t+\delta)+\wp(z(x)+\mu t)+\wp(\delta))}\right)^2}{2t}}}{\sqrt{2t}}\left[\sqrt{\kappa}\frac{\left(\wp(z(x)+\mu t)-\overbrace{\wp(+\delta)}^{e_3}\right)\left(\frac{dzdx}{dxdt}+\mu\right)}{\sqrt{2t}} - \kappa\frac{\left(\wp(z(x)+\mu t+\delta)+\wp(z(x)+\mu t)+\overbrace{\wp(\delta)}^{e_3}\right)}{2\sqrt{2}\sqrt{t^3}}\right] \quad (32)$$

Using $\sqrt{(\wp(z(x)+\delta)+\wp(z(x))+\wp(\delta))} = \varsigma(x+\delta)-\zeta(x)-\zeta(\delta)$ and $\frac{1}{2}\frac{\wp\prime(z)-\wp\prime(y)}{\wp(z)-\wp(y)} = \varsigma(z+y)-\zeta(z)-\zeta(y)$ (32.1) [12] p.451

and $\frac{dx}{dt} = \frac{1}{\sqrt{2^n n!}}\mathrm{erf}\,\frac{\sqrt{\kappa(\wp(z(x)+\mu t+\delta)+\wp(z(x)+\mu t)+\wp(\delta))}}{\sqrt{2t}}$. The differential $\frac{dz}{dx}$ may be calculated using Eq. (26) resp. Eq. (27) for resp. the upper and lower airfoil surfaces. Eq. (32.1) is relevant to decipher the related optimized p-function with translated argument.

The shear stresses may be calculated with Eq. (33) and yields with $c_1 < 0$ ($c_1$ *as in* Eq. 23).,

$$\overrightarrow{\tau_\omega} = \nu\frac{d\overrightarrow{u^*}}{dx} = \nu\frac{d}{dx}\frac{1}{\sqrt{2^n n!}}\mathrm{erf}\,\frac{\sqrt{\kappa\big(\wp(z(x)+\mu t+\delta)+\wp(z(x)+\mu t)+\wp(\delta)\big)}}{\sqrt{2t}} =$$

$$\frac{1}{\sqrt{2^n n!}}2i\nu\sqrt{12}\sqrt{\frac{L}{R_e}}\frac{\sqrt{\frac{2}{\pi}}e^{-\left(\sqrt{\kappa(\wp(z(x)+\mu t+\delta)+\wp(z(x)+\mu t)+\wp(\delta))}\right)^2/2t}}{\sqrt{t}}\left(-e_2+\wp(z(x)+\mu t)\right)\frac{dz(x)}{dx} \quad (33)$$ ), with $\nu$ the dynamic viscosity.

Using a Sagemath code to calculate the airfoil section results (listed in table 1 and) using Eqs. (32-33) yields (Please see also fig. 16-18 for an illustration of the calculated iso-pressure lines for inflow velocity varying between 10m/s to 100 m/s and the length scale in Eq. (29) set to 10 m together with the visualization of the airfoilcontour into the activity sample domain. E.g. in the illustration, an iso-pressure line of -25, has to be interpreted and corrected with the used length-scale and then added to the atmospheric pressure of 101,325 Pa to calculate the absolute pressure; The time-scale has also to be corrected with used length-scale:

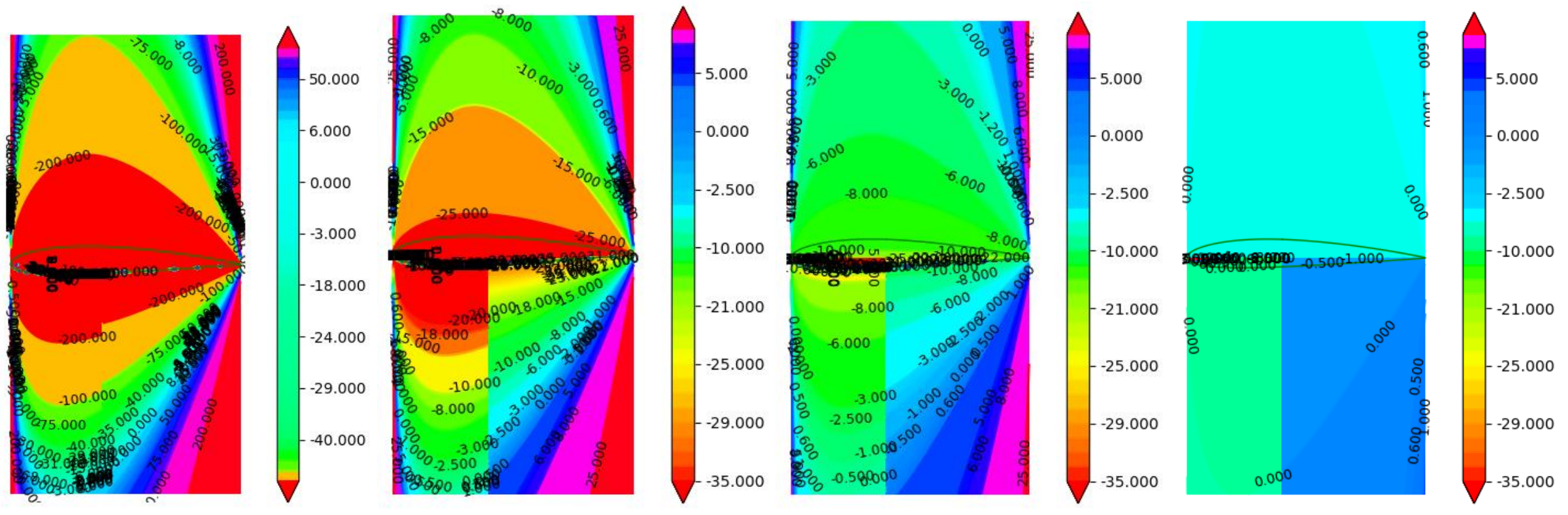

(a) (b) (c) (d)

FIGURE 16 (a)-(d) relative PRESSURE STRESS DISTRIBUTION $\frac{p_r}{v_0}$ ON A NACA AIRFOIL 2412 FOR INFLOW VELOCITY $v_0$ =49M/S, $\frac{T}{L}$=0.001 S, $\frac{T}{L}$=0.005 S resp. at $\frac{T}{L}$ =0.01 s for L=10M(RE ~3.2x10^6). The wave front of the solitary wave is then at $\frac{T}{L} * v_0$*l distance from the origin compared to the chord length of 1 m.

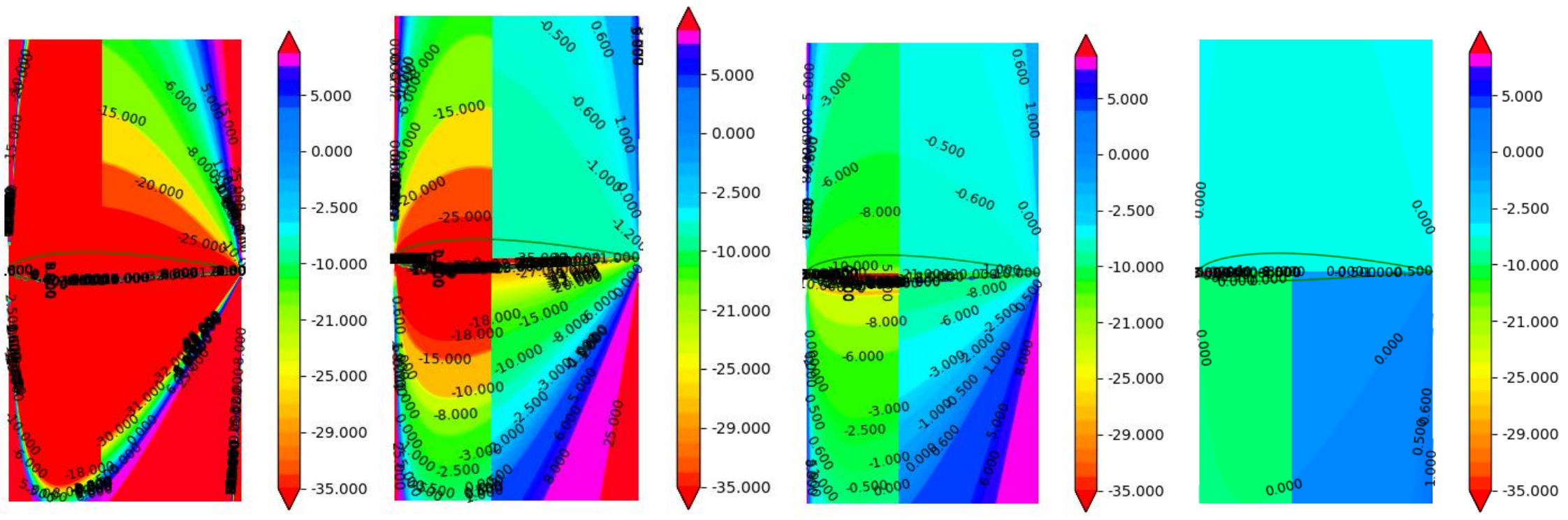

(a) (b) (c) (d)

FIGURE 17 (a)-(d) Relative pressure stress distribution $p_r{}^* = \frac{p_r}{v_0}$ ON A NACA airfoil 2412 for inflow velocity $v_0$=10M/S, $\frac{T}{L}$=0.001S , $\frac{T}{L}$=0.005 S RESP. $\frac{T}{L}$ =0.01 S and $\frac{T}{L}$=0.1 s FOR L=10M(RE ~6.6x10^5). The wave front of the solitary wave is then at $\frac{T}{L} * v_0$*L distance from the origin compared to the chord length of 1 m.

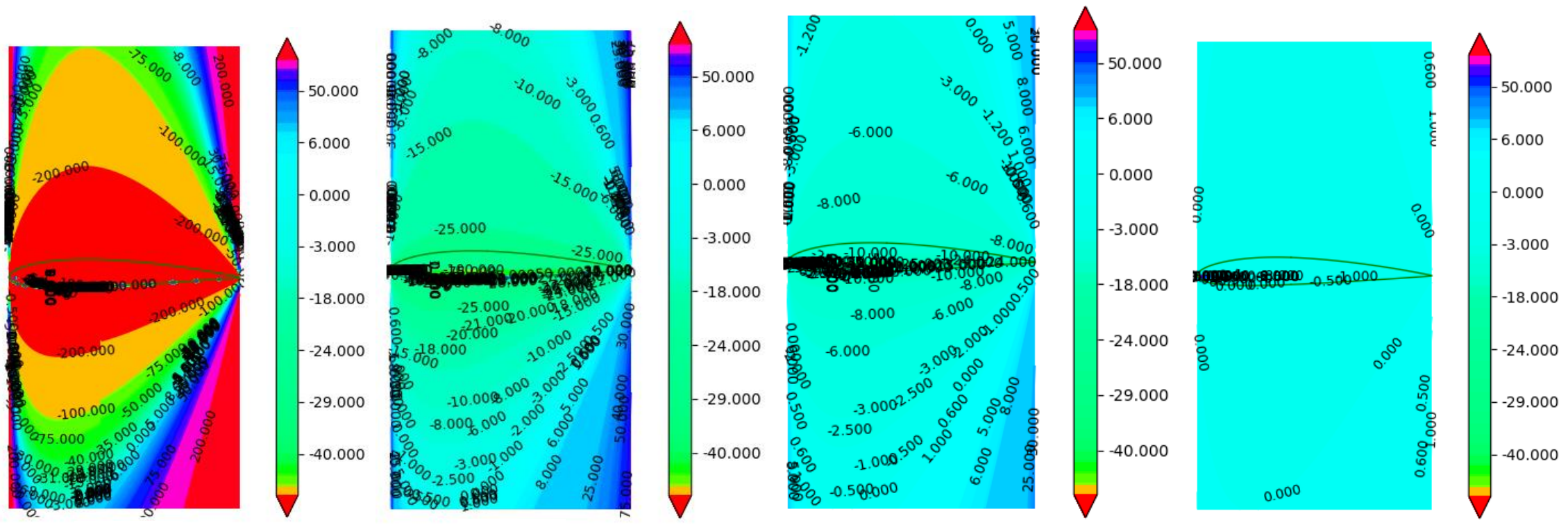

(a) (b) (c) (b)

FIGURE 18 (a)-(d) relative PRESSURE STRESS DISTRIBUTION $\frac{p_r}{v_0}$ ON A NACA AIRFOIL 2412 FOR INFLOW VELOCITY $\boldsymbol{v_0}$ =100M/S, for resp $\frac{T}{L}$=0.001 S, $\frac{T}{L}$=0.005 S, $\frac{T}{L}$ =0.01 s and $\frac{T}{L}$=0.1 s for L=10M (RE ~3.2x10^7). The wave front of the solitary wave is then at $\frac{T}{L}$* $\boldsymbol{v_0}$*l distance from the origin compared to the chord length of 1 m.

The chord lengh for the NACA 2412 airfoil is set to 1. The iso-pressure lines are characterized by smooth transitions with regard to the airfoil sections whence the inflow velocity is congruential with the design-built operating velocity range for the airfoil. At lower speeds like10 m/s are the discontinuities within the iso-pressure lines clearly visible in the lower section resp. upper section of the studied airfoil. This would imply that both the pressure distribution and the lift distribution will have discontinuities for low operating velocities which may cause aerodynamic stall. The adviced cruising speed for the Cesna 172 R aircraft holding an NACA 2412 airfoil is 63 m/s (Please see also fig 17 A-C as an example for the discontinued iso-pressure lines).

Impression of the shear stresses around the NACA 4212 airfoil are depictured in below Fig. 19 (a) - Fig. 19 (e) at time t between 0.001 s and 0.1s and $\boldsymbol{v_0} = 10\ m/s$ , chordlength c =1, $\nu = 1.516.\,10^{-5}\,\frac{m^2}{s}$, the kinematic viscosity of air at a temperature of $20^{\circ}$Celcius and Reynoldsnumber $\mathrm{R_e} = \frac{v_0 c}{\nu} \equiv \mathrm{R_e} = \frac{v_0 c}{\nu} = \frac{10.1}{1.516.10^{-5}} \equiv 659630.6$. The figures show the forming of a boundary layer around the airfoil during the measuring time. Contrarily to pressure stresses are shear stresses only defined, and thus will have only a significant physical value, on the airfoil surface or on the formed boundary layer surfaces. The Sage codes are available upon request.

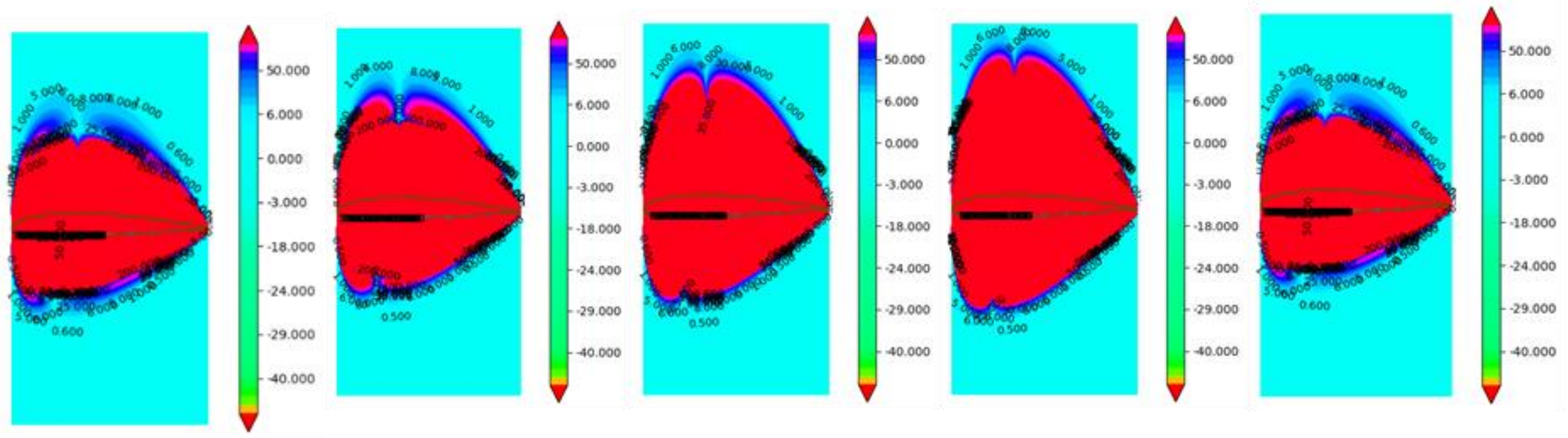

(a) (b) (c) (d) (e)

FIGURE 19 (a)-(e) Shear stresses $\boldsymbol{\tau_\omega}$ **in x − direction** around an NACA airfoil 2412 facing an inflow wind velocity of 10 m/s at Re=659630.6 . multimedia view and some frames of boundary layer formation of the Naca 2412 airfoil simulation of $\boldsymbol{\tau_\omega}$ **in x − direction** between $\frac{T}{L}$=0seconds and $\frac{T}{L}$=0.1 seconds and a growing boundary layer is observed. Contrarily to the pressure stresses around an airfoil, are shear stresses only defined on the airfoil surface or on the airfoil boundary layer surface.

Please see Fig. 20 (a) for multimediaview and some frames Fig. 20 (b) – Fig. 20 (e) of the boundary layer formation and separation ($\tau_{\omega} = \nu \frac{\partial u}{\partial y}$) shear stress analysis in y-direction for the NACA 2412 Airfoil at $\boldsymbol{v_0}$=10m/s simulation. The boundary layer separates from the airfoil whence $\frac{\partial u}{\partial y} = 0$, is attached whence $\frac{\partial u}{\partial y} > 0$ and detached whence $\frac{\partial u}{\partial y} < 0$.[60][61]

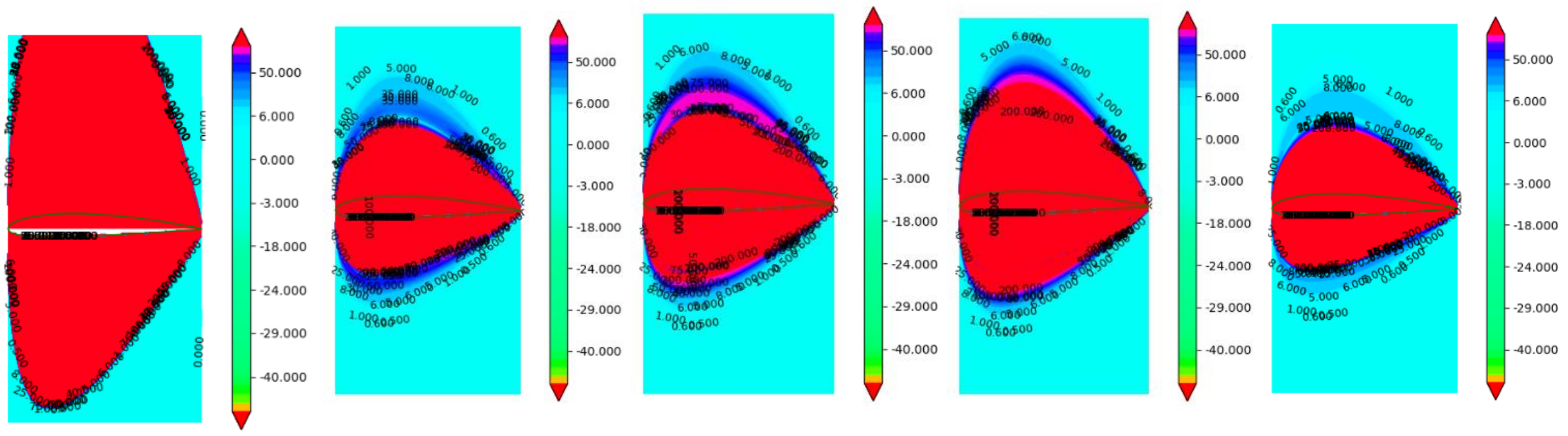

(a) (b) (c) (d) (e)

FIGURE 20 (a) multimedia view and some frames (b)-(e) of boundary layer formation and separation of the Naca 2412 airfoil at $\boldsymbol{v_0}$=10m/s simulation $\boldsymbol{\tau_{\omega}}$ **in y − direction** between $\frac{T}{L}$=0seconds and $\frac{T}{L}$=0.1 seconds

Fig. 21 (a)- Fig. 21 (f) do illustrate the still-frames of the real-values of shear stresses $\boldsymbol{\tau_{\omega}}$ in -y direction around a NCA 2412 airfoil between $\frac{t}{L}$=0 seconds and $\frac{t}{L}$= 0.1 seconds and Fig. 21 (a) is a multi-media view of this process. It is remarkable to observe that the flow separation process is a dynamic and volatile process. Please do see the multi-media view in fig 21. (g)

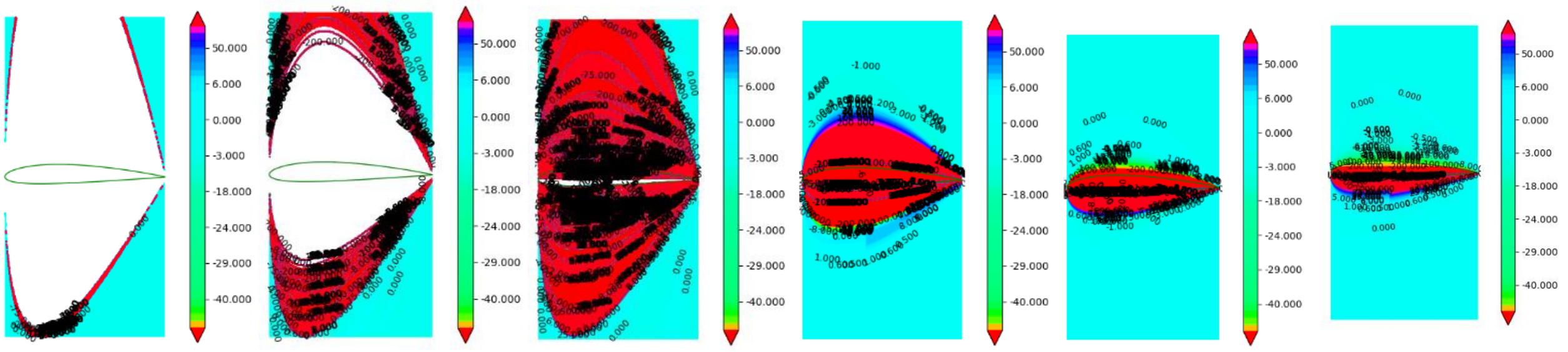

(a) (b) (c) (d) (e) (f)

FIGURE 21 (a)-(f) some stillframes for the dynamic process of boundary layer formation and separation of the Naca 2412 airfoil at $\boldsymbol{v_0}$=100m/s simulation $\boldsymbol{\tau_{\omega}}$ **in y − direction** for $\frac{T}{L}$=0.001, 0.005, 0.01 seconds and $\frac{T}{L}$=0.1 ,0.13 and 0.15 seconds at Re=6596306 (6.6$\mathbf{10^6}$). At $\frac{T}{L}$=0.15s is the flow completed separated on the upper side of the airfoil

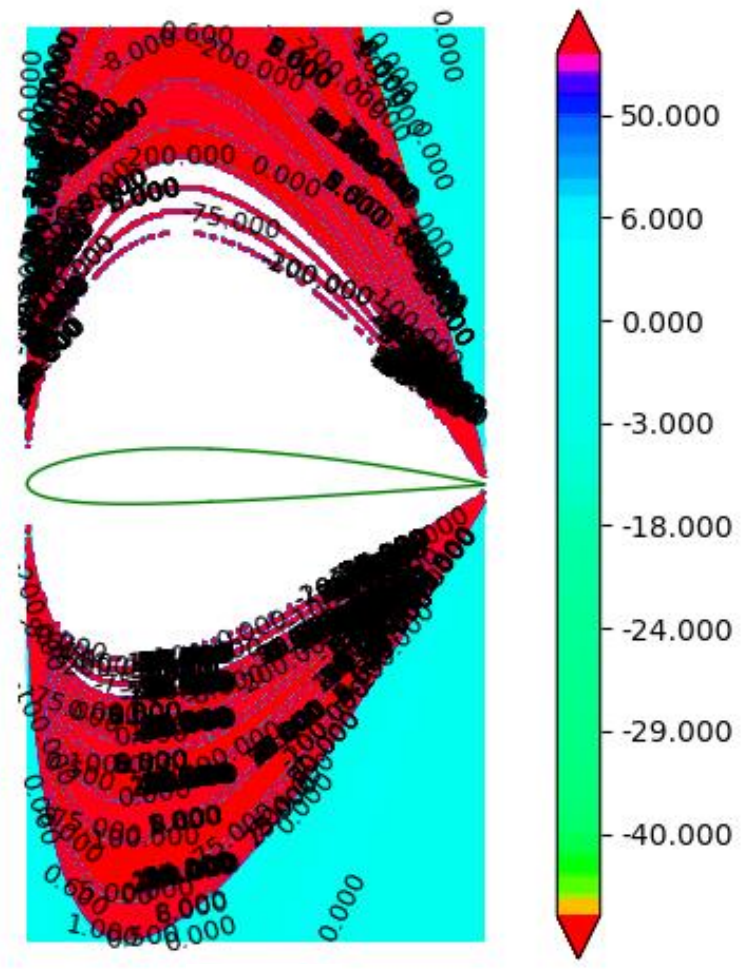

(g)

FIGURE 21(g) Multi-media view for the dynamic process of boundary layer formation and separation of the Naca 2412 airfoil at $\boldsymbol{v_0}$=100m/s simulation $\boldsymbol{\tau_\omega}$ **in y − direction** for $\frac{T}{L}$ BETWEEN 0.001 seconds and $\frac{T}{L}$=0.151 seconds at Re=6596306 (6.6$\mathbf{10^6}$). The shear stresses do revolt around the aerofolic element as the motion of the poles would predict as in Fig. 9 and Fig. 10.

## IV. UNIT CIRCLE MAPPING TO EMULATE THE OF ANGLE OF ATTACK EFFECTS WITH REGARD TO THE USED COORDINATE PARAMETRIZATIONS (VARYING THE PITCHING DERGEE OF THE NACA AIRFOIL 4 SERIES TO CALCULATE THE LIFT AND THE DRAG FORCES)

Whence the chord-length c of an airfoil is mapped with an auxiliary (unit) circle, as c=1, the coordinates ($x_n$, $y_n$) on the surface of the airfoil are then mapped with coordinates (x,y) on the (unit) circle through the following formula with

$$x = \tfrac{1}{2}(1-\cos\theta) \tag{34}$$

and

$$\theta = \operatorname{atanh}\frac{dy_c}{dx} \tag{35}$$

and $y_c$ the definition of the camber line. [17]. Using Eqs (34) & Eqs. (35) yields

$$\begin{cases} x^*{}_n = x_{U,L}\left(\overbrace{\frac{1}{2}}^{c}\left(1-\left[\overbrace{\underbrace{(1-2x)}_{\cos\theta}\,\underbrace{\cos\alpha}_{\cos\alpha}-\underbrace{\sqrt{1-(1-2x)^2}}_{\sin\theta}\left(\underbrace{\sin\alpha}_{\sin\alpha}\right)}^{\cos(\theta+\alpha)}\right]\right)\right) \\ y^*{}_n = y_{U,L}\left(\frac{1}{2}\sin\alpha\sqrt{1-(1-2x)^2}+\left(x-\frac{1}{2}\right)\cos\alpha+\frac{1}{2}\right) \end{cases} \tag{36}$$

The angle of rotation [20] may be understood by the following conformal transormations (the projection on the auxiliary circle and translation over α radians and then inverse projection on the translated airfoil): $P_\theta \circ R_\alpha \circ P^{-1}{}_{\theta+\alpha}$ $(x_n,y_n) \overset{P_\theta}{\rightarrow} (x,y) \overset{R_\alpha}{\rightarrow} (x',y') \overset{P^{-1}_{\theta+\alpha}}{\rightarrow} (x^*_n, y^*_n)$. Please do see also Fig. 23 for the visual details. The red airfoil with 0 °radians A-o-A is transformed (rotated) into the blue one with α as angle of attack and α+ θ the new camber line slope connecting (x=c/2,0) and $(x^*_n, y^*_n)$, a point on the upper-surface of the blue airfoil as illustrated in fig. 23.

Generalizing the Fundamental Thin Airfoil equation [17] [16] [21] (chap. 8) [22] we have come to the following formulea for the normal component of the velocity vector [Normally some small angle approximations are used to deduct the dependency of vorticity distribution on the surfaces (or the camberline itself whence the airfoil is symmetric) of the airfoil and the angle of attack for 0≤θ≤5° radians, interpreting the surface itself as an extended flow stream-line or vortex sheet. Please do see also fig.23]:

$$\begin{cases} v_x = v_\infty \cos\alpha + v_t \cos\varphi - v_n \sin\varphi \\ v_y = v_\infty \sin\alpha + v_t \sin\varphi - v_n \cos\varphi \end{cases} \tag{37}$$

Out of Eq. (37) and

$$\frac{v_y}{v_x} = \frac{dy_c}{dx} \tag{38}$$

yields Eq. (39) with $y_c$ equivelant to the camber line definition of the Naca Airfoil 4 series, $\varphi = \alpha + \theta \wedge \frac{dy_c}{dx} = \tan(\alpha + \theta) \wedge$ using the trigonometric identities of $\cos(\operatorname{atan}(x)) = \frac{1}{\sqrt{1+x^2}}$ and $\sin(\operatorname{atan}(x))\tan(\tan(x)) = \frac{x^2}{\sqrt{1+x^2}} \wedge \overbrace{v_\infty}^{u_{tangential}} \left(\frac{dy_c}{dx}\cos\alpha - \sin\alpha\right) = \overbrace{v_n}^{u_{normal}} (\cos\varphi + \sin\varphi\tan\varphi)$ [17] [16] [21],

$$u_{normal} = u_{tangential} \cdot \left( \frac{\frac{dy_c}{dx}\cos\alpha - \sin\alpha}{\sqrt{1+\left(\frac{dy_c}{dx}\right)^2}} \right) \tag{39}$$

with α equivelant with the angle of attack and $u_{tangential}$ the stream-line velocity as solution of the underlying Calogero-Moser Hamiltonian with elliptic potental. Combining the used coordinate (conformal) mapping and the Fundamental Thin Airfoil equations yields,

$u_{normal}(x^*_n, y^*_n) = u_{normal}(x_n, y_n) \bmod \propto |\delta|$, with $|\delta|$the observed period. (Please see also fig. 22) Hence since the conformal transform, $P_\theta \circ R_\alpha \circ P^{-1}_{\theta+\alpha}$ , is angle-preserving we may write

$$P_\theta \circ R_\alpha \circ P^{-1}_{\theta+\alpha} u_{normal}(x_n, y_n) = u_{normal}(P_\theta \circ R_\alpha \circ P^{-1}_{\theta+\alpha}(x_n, y_n)) = u_{normal}(x^*_n, y^*_n) \tag{40}$$

The last equation yields

$$u_{normal}(x^*_n, y^*_n) = u_{normal}(x_n, y_n) = (\sin\alpha - \cos\alpha) \overbrace{u_{tangential}}^{\frac{1}{\sqrt{2^n n!}} \operatorname{erf} \frac{\sqrt{(\operatorname{atanh}(z(x)+\delta)+(\operatorname{atanh}(z(x))+(\operatorname{atanh}(\delta))}}{\sqrt{2t}}} \tag{41}$$

for $\theta = \frac{\pi}{2} - \alpha$ which is the blue airfoil in fig. 23 that transposes to the red (initial airfoil) for $\alpha = 0°$. Since the used transform is angle-preserving does it imply that $u_{normal}(x^*_n, y^*_n) = u_{normal}(x_n, y_n)$ for all values of θ. As a matter of fact the atanh(x) components present in the definition of the tangential velocity component distribution does represent a hyporbolic angle.

Therefor will the normal velocity and so the pressure stresses (and the Lift) oscillates in the normal direction. This can be seen as follows. Let us take a simple and not unphysical expample of $\alpha = \theta$ then is the function $\frac{u_{normal}}{u_{tangential}} = 3\,\mathrm{sign}(\sec(2\alpha))\sin(\alpha)$ a wavy discontinuous function as a function of the A-o-A α:illustated below in Fig. 22

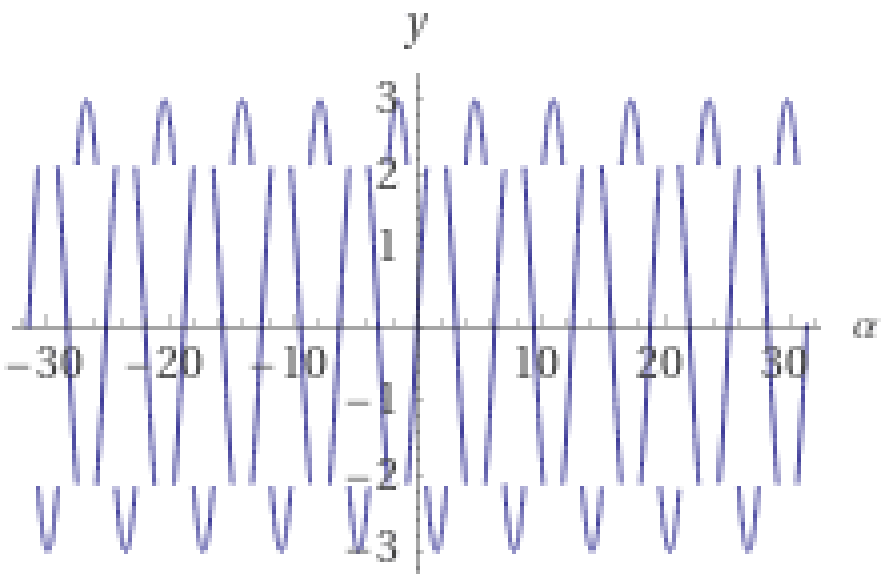

FIGURE 22 Wavy and discontinuous patterns in the normal velocity distribution of a NACA 2412 airfoil

Obviously the speed control mechanism of modern aircrafts may manage the jumpy velocity distribution through acceleration at critial (disconiuity points) to minimize the (duration and thus the) effects of flow seperation (the fall-out of lift). Whence applying the solutions of the non-dimensionalized Navier Stokes d.e. Eq. (29) to the unit circle mapping transform as depictured in fig. 23 we must also correct the arclength (= radius times the enclosed angle), and so the pitching angle or the angle of attack α as demanded by Eq. (29.3), $\alpha^* = \frac{\alpha}{L}$. A sensitivity analysis for the studied variable Lift and (induced) Drag [23] [24] on the NACA airfoil 2412 with regard to the A-o-A α is depictured in fig. 24 in the case of $v_0 = 10m/s$. The graph's period is precisely 10π (chosen length scale times max. slope camber line angle). If we shift the graph 5π to the right (to get physical interpretable data), then the angle of attack that would correspond with zero lift value is equivalent with angle-of-attack $\alpha_{L_0} = 0^\circ$degrees and the stall angle would be then approximately $13^\circ$degrees what is conformant with the literature and experimental data of this type of airfoil. [62] p.478 [63]

Out of Eq. (39) can the Lift component be calculated as follows

The following formulae are derived:

The pressure stress =$u_{\text{normal}} = \frac{m}{A}\frac{du_t}{dt}$ (42)

The shear stress = $\nu\frac{du_t}{dx}, \nu\frac{du_t}{dy}$ (43)

( the shear stess is a tensor, and may exist in all directions)

The normal force per meter span $N = \int_{LE}^{TE} p_L(x,-y)dx + \int_{LE}^{TE} p_U(x,y)dx$ [23] [24], using Eq. (32) for the found pressure distribution on an unsupported (unmounted) NACA 2412 airfoil, is then equivalent with, with the length scale chosen as 10m and $v_0 = 10m/s$ (Please see also Eq. (45) setting $\alpha = 0°$ .The codes are available upon request.)

```
sage: integral(pr_LLeft(x,0.04,0.01,1.516*10^(-5),-1.854074*I)+pr_ULeft(x,0.08,0
....: .01,1.516*10^(-5),-1.854074*I), x, 0.00001, 0.4).abs()+integral(pr_LRight(
....: x,0.04,0.01,1.516*10^(-5),-1.854074*I)+pr_URight(x,0.08,0.01,1.516*10^(-5)
....: , -1.854074*I), x, 0.4,1).abs()
3.31320919902823
```

$N = \int_0^1 p_L(x,-y,t,\mu,\delta) \times v_0\, dx + \int_0^1 p_U(x,y,t,\mu,\delta) \times v_0\, dx \approx 10 \times$3.31320919902823

≈33.1320919902823N *using* Eq.(45) without the $2\pi i$ branch surface correction of Eq. (25)

At inflow velocity of 63 m/s (which is the cruising speed of a Cessna 172 T) is the transient generated normal force per meter span per unit velocity under A-o-A = 0 ° radians) at t=0.001s ($N \approx 16.6238 \times 63 = 1047.25N$ *using* Eq.(45))

```
sage: integral(pr_LLeft(x,0.04,0.001,1.516*10^(-5),-1.854074*I)+pr_ULeft(x,0.08,
....: 0.001,1.516*10^(-5),-1.854074*I), x, 0.0001, 0.4).abs()+integral(pr_LRight
....: (x,0.04,0.001,1.516*10^(-5),-1.854074*I)+pr_URight(x,0.08,0.001,1.516*10^(
....: -5), -1.854074*I), x, 0.4,1).abs()
16.6238086383881 N without the 2πi branch surface correction of Eq. (25)
```

```
At t=0.01 s is the transient generated normal force per meter span per unit velocity under
A-o-A = 0 °radians) (N≈ 0.5111053 × 63 = 32.193N using Eq.(45))
sage: integral(pr_LLeft(x,0.04,0.01,1.516*10^(-5),-1.854074*I)+pr_ULeft(x,0.08,0
....: .01,1.516*10^(-5),-1.854074*I), x, 0.0001, 0.4).abs()+integral(pr_LRight(x
....: ,0.04,0.01,1.516*10^(-5),-1.854074*I)+pr_URight(x,0.08,0.01,1.516*10^(-5),
....:  -1.854074*I), x, 0.4,1).abs()
0.511105348885139 N without the 2πi branch surface correction of Eq. (25)
```

Recall that the dimensionless Navier Stokes d.e is as follows:

$$\frac{\mathrm{Du}}{\mathrm{Dt}} + \mathrm{u}.\nabla \mathrm{u} = -\frac{1}{\rho}\nabla \mathrm{p_r} + \frac{1}{R_e}\nabla^2 \mathrm{u} + \mathrm{F} \text{ with } \mathrm{R_e} \equiv \frac{1}{\mu}$$

The results for the total normal forces per meter span match the (stationary) NASA experimental (wind tunnel) data of airfoil 2412 of the lift variable which is 35.83N in similar circumstances with $\mathrm{v_0} = 10\frac{\mathrm{m}}{\mathrm{s}}$ and angle-of-attack 0°degree [25] [26] (side-mounted), and $|N| \equiv 34.0982$ N (unmounted) for t=0.01 seconds (whence the front of the soliton wave is at 0.1 chord length: $0.01\mathrm{s}\times 10\frac{\mathrm{m}}{\mathrm{s}} = 0.1m$)

Hence

$$\vec{N} = (+3.29028179591000 + 0.266109630423300 * I)*v_0 + \overbrace{2\pi * \mathrm{I}}^{\text{branch surface correction.Please do see Eq.(25)}} = 34.0982e^{0.265567i}N$$

```
sage: integral(-pr_LLeft(x,0.04,0.01,1.516*10^(-5),-1.854074*I)-pr_ULeft(x,0.08,
....: 0.01,1.516*10^(-5),-1.854074*I), x, 0.00001, 0.4)+integral(-pr_LRight(x,0.
....: 04,0.01,1.516*10^(-5),-1.854074*I)-pr_URight(x,0.08,0.01,1.516*10^(-5), -1
....: .854074*I), x, 0.4,1)
3.29028179591000 + 0.266109630423300*I
```

The observed and reported lift in the studied literature is likely a time-average one or stationery one. The complex component of the Lift variable will damp faster in time than the corresponding real component. Usually, transient models are replaced by stationery ones due their simplicity and comparable results for non-turbulent studies concerning the A-o-A measuring smaller than 8°degrees. [6]

Taking the effect of the A-o-A into consideration, using Eq.(32) & Eq. (33), can the Drag resp.Lift [23] [24] be represented as (whence the airfoil is side-mounted and can rotate only on its chord length axis as shown in Fig. 23), with T a "quick" damping tensor,

$$\mathrm{D} = -(\cos\alpha\, T - \sin\alpha\, N) = -\int_{x_{LE}}^{x_{TE}} \left( \nu \frac{du_t}{dx}\cos\alpha - \sin\alpha \left( \frac{\frac{dy_c}{dx}\cos\alpha}{\sqrt{1+\left(\frac{dy_c}{dx}\right)^2}} - \sin\alpha \right) \frac{m}{A}\frac{\partial}{\partial t}u_t \right) dx \quad (44)$$

$$\mathrm{L} = \mathrm{N}\cos\alpha - \mathrm{T}\sin\alpha = \int_{x_{LE}}^{x_{TE}} \left( -\nu \frac{du_t}{dx}\sin\alpha + \cos\alpha \left( \frac{\frac{dy_c}{dx}\cos\alpha}{\sqrt{1+\left(\frac{dy_c}{dx}\right)^2}} - \sin\alpha \right) \frac{m}{A}\frac{\partial}{\partial t}u_t \right) dx \quad (45)$$

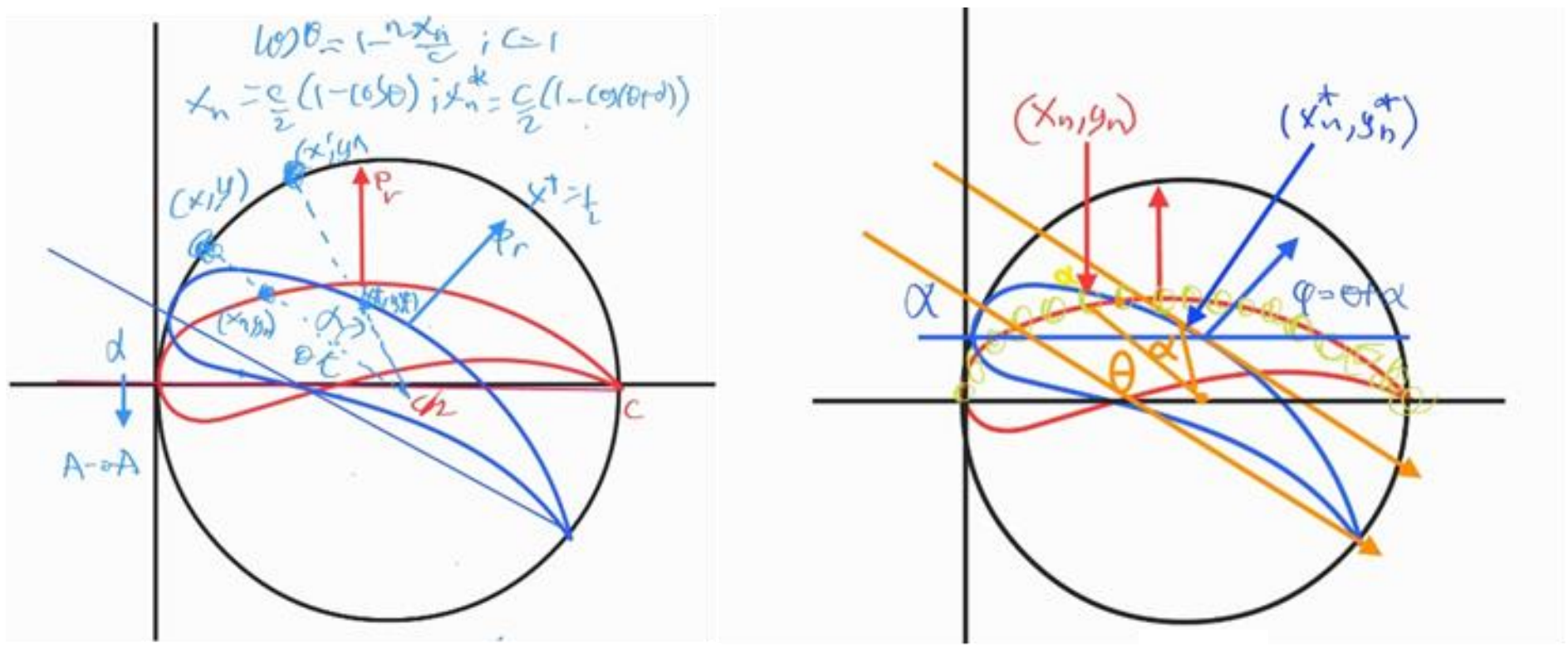

(a) (b)

FIGURE 23 (a)-(b) Angle-of-Attack emulation through a conformal coordinate transformation, a Vortex sheet present on the upper surface of the right graph. The last is interpreted as an extension of the stream-lines in the Thin Airfoil theory.

The stall angle which is the angle, that is independent of the chord length variable x, where the lift forces do become null may be read out of the third illustration in below Fig. 24 and is around 13 degrees for the studied airfoil which is conformant with the studied literaure and experimental data as noticed before.

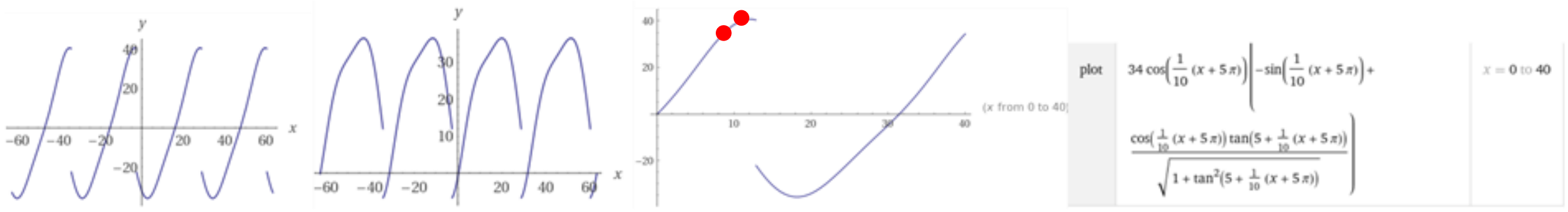

(a) (b) (c) (d)

FIGURE 24 (a)-(d) Sensitivity analysis for The Lift and (induced,T$\approx \mathbf{0}$) Drag [23] [24] variables with regard to the angle of attack $\boldsymbol{\alpha}^* = \frac{\boldsymbol{\alpha}}{L}$ (and θ=5 as in eqs 44 & 45) for the NACA airfoil 2412 at $\mathbf{v_0} = \mathbf{10}\,\frac{\mathbf{m}}{\mathbf{s}}$, length scale l=10m, and Re=659630.6 using eqs. (44) resp. (45) at t=0.01 s. In the snapshots of the recurring cycles of the studied lift and drag variables are the predicted distribution discontinuities clearly visible for varying pitching angles of the angle-of-attack. The third illustration is the graph of the measured lift as a function of the angle-of-attack shifted 5π radians to the left. The stall angle is clearly visible. The outer right illustration is the mathematical expression for the lift with A-o-A as variable x and θ=5 radians, shifted 5π radians.

## V. DISCUSSION OF RESULTS

The tangential velocity distribution is calculated through the Navier Stokes d.e. The normal velocity distribution is also a solution of the Navier Stokes d.e. aiding with the Thin Airfoil theory for the boundary layer regions, without the usual industry approximations.

The Lift (and the Drag) forces on an airfoil are critically depended on the ratio of their mass and (contact) surface area and the Reynoldsnumber related to the flow velocity.

Normally is that ratio non-uniformingly distributed across the airfoil that need in practice to be considered (which is obviously done) since it may hold critical thresholds with regard to the corresponding Lift and Drag forces magnitude. E.g. the engine within the turbine hull makes the leading edge of an airfoil heavier than its trailing edge.

The variable time plays also a critical role with regard to the damping of the singularity-incentived aerodynamic forces at the leading edge of the airfoil due their geometrical definitions and needs further future studies.

The maximum absolute value of the measured normal forces are congruential with the unleveraged magnitude of the lift forces (the first and lower red dot in Fig. 24c). Hence under the effect of the varying pitching angle-of-attack, the maximum value of the lift forces will exceed its original and native value. This follows automatically out of the Eq. (45) and the additional begin conditions (e.g. the lift at zero degree angle of attack has to be zero, the angle-of-attack is independent of the chord line.) This new optimum value for the lift will occur moments before the studied aerofolic body reaches its stall angle (Please do see the second red dot in Fig. 24c).

The results may serve as a manual to analytically calculate pressure and shear stresses of the NACA 4 series airfoils and to better understand the flow separation that typically arises around the NACA 4 series. It is not a matter of when does the flow separation takes place but where does it takes place, which does coincide with the location with minimum absolute value of the shear and pressure stresses.

The (shear) stresses have been observed to revolt heavily around the studied aerofolic element most likely following the motion of the poles of the underlying Hamiltonian as depictured in fig. 9 and fig. 10 while considering the heavy fluctuations in their respective graphs.

The point zero (x=0) is a removable singularity for the all the velocity derivatives including the pressure and shear stress tensors due to its geometric definition caused by the radical within the definition of the thickness of the NACA 4 series airfoil. The respective graphs do consequently fluctuate heavily when the lower limit is followed to the smallest convergence point near the origin of the used (complex) Euclidean axis system.

The normal component of the velocity vector demonstrated also its volatile and oscillatory character. That should be not surprising since the begin condition, which is a quartic and does also manifest some initial fluctuations at t=0.

There are some dimensionless correlations between flow variables deducted as a consequence of the underlying conservation laws and application of the Navier Stokes d.e.

Hence, the sum of the half-periods of the used p-function in our emulations was equivalent with $\frac{\left(-\frac{i}{8}\right)\Gamma^2\left(\frac{1}{4}\right)}{2\sqrt{\pi}}$, which is a necessary condition for an exact solution to the problem. Therefor to avoid unnecessary instabilities caused by geometrical definitions of the airfoil design, we did set δ in Eqs. (30)- (33) equivalent to the sum of the half-periods.

$$\delta = -\frac{\left(\frac{i}{8}\right)\Gamma^2\left(\frac{1}{4}\right)}{4\sqrt{\pi}} \approx -1.854074i. \qquad (46)$$

The other deduction, Eq. (23.1, 3rd eq.), is about the prediction of the change of kinematic viscosity as function of the time as a consequence of the application of the continuity condition on the interior of the sample region. Both deductions are non-heuristic and analytic based.

## VI. CONCLUSION

The presented innovative analytical model for calculating the shear and pressure stresses does give us surprisingly more insight in its computational capabilities as a power tool that may be used for fast-designing efficient airfoils. Discontinuity “holes” in the pressure and shear stresses distributions of studied wing elements caused by geometric design faults are then simple calculated and visualized.

The analytical results do correspond with the numerical and experimental benchmark data gathered for up to certain accuracy (our model is transient in contrast to the general used stationary models with regard to lift computations on wing sections). Wind tunnels results and thus most benchmark studies do consider a base-supported and stationery airfoil where some (transient) forces are seemingly neutralized or neglected. The flow problem over an airfoil is equivalent to one related to the flow within a driven-lid cavity. Both problems yield a degenerate form of the Weierstrass p-function as begin condition potential, expressed as the square of the inverse hyperbolic tangent function.

Singular points around an airfoil which arises due the geometrical definitions may be critical and are easy identifiable and made visible. The presented method is analytically intensive but both a computer- and financial efficient way to calculate the pressure and shear stresses' distributions around an airfoil (or any mathematically modelled object) that were until now only computable via wind tunnels armed with watertube-filled manometers or via turbulence models with empirical data from the wind-tunnels setup models.

The height (and width) of the fins of the saltwater hunting fishes tend to assume an optimized magnitude with regard to the maximum amplitude of the stress forces, to optimize their maneuverability as much as possible and to contra-arrest regions where flow separation may occur due to excessive flow velocity.

The theoretical concept of stagnation point with regard to an airfoil is contained within the realm of the found 1-soliton solution of the velocity distribution for t>0. Hence $\lim_{x\to 0} \vec{u}(\xi(x,y),t) = \lim_{x\to 0} \frac{u_\infty}{\sqrt{2^m m!}} \operatorname{erf} \frac{\sqrt{\wp(\xi+\delta)}}{\sqrt{2t}} = \operatorname{erf}(\operatorname{atanh} 0) = 0$ whence $\delta$ is chosen as the negative sum of the half-periods ($\omega_3 = -\omega_2 - \omega_1$) of the related p-function, with $\wp(\omega_3) = -\wp(-\omega_2 - \omega_1) = e_3 = 0$ in the case of a NACA 4 series airfoil and $\omega_2 + \omega_1 = i\, 1.854074$.

The choice of the parameter $\delta$ determines the volatility of the tensor stresses such as the pressure and shear stresses.

Flow separation may also occur due pitching beyond the stall angle, i.e. increment of the angle-of-attack beyond a threshold where flow separation becomes eminent due to insufficient or negative lift forces and unproportionally increasing drag forces. Boundary layer forming around an airfoil, what critical is for the wing element's stability, may be made visible with the found transient distribution of the related shear pressure stresses.

The presented analytical results may easily be extended to the other NACA airfoil series or any other analytically defined airfoil, obviously with different nuances in shear and pressure stresses distributions due their typical geometric definitions.

## VII. RECOMMENDATIONS FOR FOLLOW-UP STUDIES

The variable time plays also a critical role with regard to the damping of the singularity-incentived Drag forces near the edges of the NACA 4 series airfoil due their geometrical definitions and needs further studies in transient models.

Not at least is also recommendable the study with regard to the found correlation between the area and weight of the Naca series airfoils as condition to its flight stability.

## DATA AVAILABILITY

The data that support the findings of this study are available within this paper or its references. The author is willing to supply additional data upon reasonable request.

## CONFLICT OF INTEREST STATEMENT

There is no conflict of interest.

## FUNDING DECLARATION

There are no funding agencies involved in the production of this manuscript.